\documentclass[manuscript]{revtex4}

\usepackage{amsfonts}
\usepackage{mathrsfs}
\usepackage{amsmath}
\usepackage{amssymb}
\usepackage{bm}
\usepackage{feynmp-auto}
\usepackage[normalem]{ulem}
\usepackage{extarrows}
\usepackage{slashed}
\usepackage{isodateo}
\usepackage{graphicx}
\usepackage{xcolor}
\usepackage[CJKbookmarks=true, bookmarksnumbered=true,bookmarksopen=true]{hyperref}
\hypersetup{colorlinks,%
	linkcolor=[rgb]{0,0.2,0.6}, %
	urlcolor=[rgb]{0,0.2,0.6}, %
	citecolor=[rgb]{0,0.2,0.6}}

\usepackage{indentfirst}

\graphicspath{{fig/}}

\usepackage{tikz}
\usetikzlibrary{arrows,shapes}
\usetikzlibrary{trees}
\usetikzlibrary{matrix,arrows} 				
\usetikzlibrary{positioning}				
\usetikzlibrary{calc,through}				
\usetikzlibrary{decorations.pathreplacing}  
\usepackage{pgffor}							

\usetikzlibrary{decorations.pathmorphing}	
\usetikzlibrary{decorations.markings}
\tikzset{
	photon/.style={decorate, decoration={snake}, draw=red},
	electron/.style={draw=blue, postaction={decorate},
		decoration={markings,mark=at position .55 with {\arrow[draw=blue]{>}}}},
	gluon/.style={decorate, draw=black,
		decoration={coil,amplitude=4pt, segment length=4pt}} ,
	vector/.style={decorate, decoration={snake}, draw},
	provector/.style={decorate, decoration={snake,amplitude=2.5pt}, draw},
	antivector/.style={decorate, decoration={snake,amplitude=-2.5pt}, draw},
	fermion/.style={draw=black, postaction={decorate},
		decoration={markings,mark=at position .55 with {\arrow[draw=black]{>}}}},
	fermionbar/.style={draw=black, postaction={decorate},
		decoration={markings,mark=at position .55 with {\arrow[draw=black]{<}}}},
	fermionnoarrow/.style={draw=black},
	fermionnoarrowsoft/.style={draw=blue},
	scalar/.style={dashed,draw=black, postaction={decorate},
		decoration={markings,mark=at position .55 with {\arrow[draw=black]{>}}}},
	scalarbar/.style={dashed,draw=black, postaction={decorate},
		decoration={markings,mark=at position .55 with {\arrow[draw=black]{<}}}},
	scalarnoarrow/.style={dashed,draw=black},
	scalarnoarrowsoft/.style={dashed,draw=blue},
	electron/.style={draw=black, postaction={decorate},
		decoration={markings,mark=at position .55 with {\arrow[draw=black]{>}}}},
	bigvector/.style={decorate, decoration={snake,amplitude=4pt}, draw},
}

\tikzstyle{block} = [draw, rectangle, 
minimum height=3em, minimum width=6em]
\usetikzlibrary{arrows,patterns}

\begin{document}

	\title{Classical Limit of Yukawa theory from quantum state perspective}
	
	\author{Kaixun Tu}
	\email{tkx19@tsinghua.org.cn}
	\affiliation{Department of Physics, Tsinghua University, Beijing 100084, China 
	}
	\author{Qi Chen}
	\email{chenq20@mails.tsinghua.edu.cn}
	\affiliation{Department of Physics, Tsinghua University, Beijing 100084, China 
	}
	\author{Qing Wang}
	\email{wangq@mail.tsinghua.edu.cn}
	\affiliation{Department of Physics, Tsinghua University, Beijing 100084, China 
	}
	\affiliation{Center for High Energy Physics, Tsinghua University, Beijing 100084, China 
	}

	\begin{abstract} 
			 We derive the quantum states corresponding to classical scalar fields in the representation expanded by the eigenstates of quantum field operators.
			  This allows us to directly observe the spatial entanglement structure of quantum states and explore the differences and relationships between quantum superposition and classical superposition.
We find that if two classical fields are identical in a certain spatial region, then their corresponding quantum states have the same reduced density matrix in that region. This indicates that knowing the classical field in a local region is sufficient to derive the reduced density matrix for that region.					
		According to the correspondence between classical quantities and quantum states, we derive the equation of motion of the classical theory from the evolution of quantum states in Yukawa theory. This leads to the relativistic classical Yukawa theory, and we further obtain the relativistic corrections to the Yukawa potential.
	\end{abstract}

	\maketitle

\section{Introduction}

The classical limit of quantum theory has been a longstanding problem since the birth of quantum mechanics.  Several well-known procedures have been developed to construct the classical theory from a given quantum theory. These methods can be broadly classified into two main categories. The first category is the $\hbar \to 0$ scenario, initially proposed by Planck in his investigation of the low-frequency and high-temperature asymptotic behavior of blackbody radiation \cite{Planck1913}. In this scenario, as Planck's constant $\hbar$ approaches zero, the energy spectrum converges to the classical equipartition theorem. The second category is the $N \to \infty$ scenario, initially proposed by Bohr in his correspondence principle \cite{Bohr1920a,Bohr1922,J.Rud.Nielsen1976}. In this scenario, the classical theory emerges from the quantum theory as the quantum number $N$ of the system tends to infinity. 

With the continuous advancement of quantum mechanics, $\hbar\to 0$ not only transforms Planck's formula for blackbody radiation into the Rayleigh-Jeans formula, but also turns the Schrodinger equation into the Hamilton-Jacobi equation and leads to the principle of least action from Feynman's path integral \cite{Dirac1981,Derbes1996,Dente1978,Landsman2005,Maslov2001}. However, most of these methods for deriving classical limits only exhibit the mathematical correspondences without revealing the physical correspondences between quantum states and classical point particles.
Actually, quantum theory possesses a greater number of degrees of freedom compared to classical theory and can encompass more intricate physical phenomena. This implies that not all quantum states have classical correspondences even in the limit $\hbar\to 0$ \cite{Rosen1964,Cohn1972a,Kazandjian2006,Kazandjian2007,Klein2012}.
For example, a physical system described by a wave function with two peaks that are far apart cannot be described by classical theory.
  Schrodinger discovered that coherent states of a harmonic oscillator serve as ideal quantum states corresponding to point particles \cite{Schroedinger1926,Howard1987}. Coherent states not only minimize the effects of the Heisenberg uncertainty principle but also maintain locality in both position and momentum during their evolution. By utilizing coherent states, the concept of point particles emerges and the classical limit of quantum theory can be attained \cite{Hepp:1974vg,Zhang1990}. 

Deriving the classical limit of quantum field theory is more complex, because the classical description of a physical system involves not only point particles but also classical fields.
 R. Glauber expanded on Schrodinger's ideas and found that coherent states in quantum field theory correspond to classical fields in classical theory \cite{Glauber1963,Glauber2006a,Klauder1968,Zhang1990}. However, coherent states in quantum field theory are typically presented in Fock space, which makes it challenging to directly show the relations between quantum states and classical fields. To provide a more direct explanation of the structures of coherent states, we use Yukawa theory as a model and derive coherent states in the representation expanded by the eigenstates of quantum field operators. 
We outline some natural requirements for quantum states that have classical correspondence.
By analyzing the structure of the vacuum state and combining the requirements, we can find the quantum states that correspond to the classical fields.
By evaluating these wave functions which turn out to be coherent states, we can derive the classical Hamiltonian equations of motion from the evolution of quantum states.
Through the explicit expression for coherent states in the representation expanded by the eigenstates of quantum field operators, we not only directly observe the spatial entanglement structure of quantum states, but also identify the differences between quantum superposition and classical superposition.

	Using the representation expanded by the eigenstates of quantum field operators, we also found that if two classical fields are identical in a certain region, then their corresponding quantum states have the same reduced density matrix in that region. This indicates that, although the characteristic of classical fields having definite values at each spatial point seems to contradict the spatial entanglement of quantum states, we can still obtain a consistent correspondence between classical fields and quantum states.	
	This also leads to a corollary: by only knowing the classical field in a local region without needing the global classical field, we can immediately write down the quantum reduced density matrix for that region.
	Of course, this does not mean that spatial entanglement plays no role.
	When two quantum states with the same reduced density matrix in a certain region are superposed, due to spatial entanglement, the superposition state no longer retains the original reduced density matrix in that region.

Moreover, our work can provide insights for deriving the classical limit of quantum gravity. Currently, quantum gravity remains beyond the reach of experimental testing due to technological limitations. Consequently, establishing whether the classical limit of quantum gravity corresponds to general relativity becomes a crucial test for the validity of the theory \cite{Smolin1996,Han2007}. In loop quantum gravity, gravity has been quantized using an approach analogous to lattice quantum field theory, and the quantum states are constructed based on the spin network basis which is defined in real spacetime coordinates rather than the traditional Fock space with creation and annihilation operators \cite{Rovelli2004, Rovelli2014, Ashtekar2021}. 
Similarly, throughout our entire paper, we have treated quantum states in the real spacetime representation, following the canonical framework of quantum field theory. 
While classical limits of some quantum gravity theories defined by path integrals (spinfoam model) have been derived under certain conditions \cite{Han2017, Han2020}, substantial progress in the classical limits of canonical quantum gravity (spin network model) is yet to be achieved.

Through the correspondence between classical quantities (point particles and classical fields) and quantum states, we obtain a fully relativistic classical Yukawa theory, which can describe the interactions of Yukawa-type charged particles traveling at extremely high speeds.
In classical electrodynamics, when the velocities of charged particles are much lower than the speed of light, their interactions can be described by the Coulomb potential with Darwin's relativistic corrections \cite{Darwin1920}. Similarly, in relativistic classical Yukawa theory, we find that the potential between stationary Yukawa-type charged particles is precisely the traditional Yukawa potential, and we derive the relativistic corrections to the Yukawa potential.
Solving the equation of motion in relativistic classical Yukawa theory is more intricate compared to electrodynamics because the retarded potential in electrodynamics depends only on one point along the particle's trajectory, whereas in Yukawa theory, the retarded potential depends on all trajectory points within the past light cone due to the mass term of the scalar field.
Notably, when the mass of the scalar field is zero, although the Yukawa potential reduces to the Coulomb potential, the relativistic corrections to the Yukawa potential do not become the Darwin's relativistic corrections to the Coulomb potential.

This paper is organized as follows. In Sec. \ref{subsec:sys}, we derive the quantum states corresponding to classical scalar field in the representation expanded by the eigenstates of quantum field operators.
	According to the correspondence between quantum states and classical fields, we derive the equation of motion for the classical field from the evolution of quantum states in Yukawa theory. We also discuss the differences and relationships between quantum superposition and classical superposition.	
	In Sec. \ref{red}, we find that if two classical fields are identical in a certain spatial region, then their corresponding quantum states have the same reduced density matrix in that region. This indicates that, by only knowing the classical field in a local region without needing the global classical field, we can immediately write down the reduced density matrix for that region.	
 In Sec. \ref{cy}, we derive the equation of motion for a Yukawa-type charged particle, thereby obtaining the complete relativistic classical Yukawa theory.
 In Sec. \ref{sec:Yukawa}, we derive the relativistic corrections to the Yukawa potential. Section \ref{sec:dis} presents conclusions and discussions.



\section{Quantum states corresponding to classical fields}
\label{subsec:sys}

The Lagrangian of Yukawa theory in quantum field theory is given by
\begin{equation}\label{Yuklag}
	\mathcal{L}=i\hbar \bar{\psi}\gamma^{\mu}\partial_{\mu}\psi{\color{black}-}m_1c^2\bar{\psi}\psi+\frac{1}{2}\partial_{\mu}\phi\partial^{\mu}\phi-\frac{1}{2}\left(\frac{m_0c^2}{\hbar}\right)^2\phi^2-\hbar^{\frac{1}{2}}c^{\frac{3}{2}}g\phi\bar{\psi}\psi  \;  ,
\end{equation}
where $m_1$ and $m_0$ are the masses of the fermion and boson respectively, and
$\partial_0 \equiv \frac{\partial}{\partial t},\partial_i \equiv c\frac{\partial}{\partial x^i}\;,i=1,2,3$.
In classical field theory, we need two separate Lagrangians to describe the entire theory. One Lagrangian describes the motion of point-like particles, while the other describes the evolution of fields. Therefore, in order to avoid dealing with too much complexity at once, we decompose the quantum Yukawa Lagrangian \eqref{Yuklag} into two components:
\begin{equation}\label{spllagl}
	\begin{split}
		&\mathcal{W}_1=\frac{1}{2}\partial_{\mu}\phi\partial^{\mu}\phi-\frac{1}{2}\left(\frac{m_0\textcolor{black}{c^2}}{\textcolor{black}{\hbar}}\right)^2\phi^2-\textcolor{black}{\hbar^{\frac{1}{2}}c^{\frac{3}{2}}}g\phi j  \;,\\
		&\mathcal{W}_2=i\textcolor{black}{\hbar }\bar{\psi}\gamma^{\mu}\partial_{\mu}\psi-m_1\textcolor{black}{c^2}\bar{\psi}\psi-\textcolor{black}{\hbar^{\frac{1}{2}}c^{\frac{3}{2}}}g\phi\bar{\psi}\psi
		\;  ,
	\end{split}
\end{equation}
where $j$ represents a classical external source, and its relation to charged particles can be derived from quantum theory, which will be discussed later.
The Hamiltonian operators corresponding to the Lagrangians \eqref{spllagl} are as follows:
\begin{equation}\label{spllag}
	\begin{split}
		&\hat{H}_1=\int d^3x\left[\frac{1}{2}\hat\pi^2+\frac{\textcolor{black}{c^2}}{2}(\nabla\hat\phi)^2+\frac{1}{2}\left(\frac{m_0\textcolor{black}{c^2}}{\textcolor{black}{\hbar}}\right)^2\hat\phi^2+\textcolor{black}{\hbar^{\frac{1}{2}}c^{\frac{3}{2}}}g\hat\phi j\right]+\Lambda  \; ,\\
		&\hat{H}_2=-i\hbar \hat{\psi}^\dagger\gamma^0\boldsymbol{\gamma}\cdot\boldsymbol{\nabla}\hat{\psi}+m_1c^2\hat{\psi}^\dagger\gamma^0\hat{\psi}+\hbar^{\frac{1}{2}}c^{\frac{3}{2}}g\hat{\psi}^\dagger\gamma^0\hat{\psi}+\Lambda'
		\; ,
	\end{split}
\end{equation}
where $\Lambda$ and $\Lambda'$ are energy constants used to adjust the vacuum energy.
The first Hamiltonian $\hat{H}_1$ describes a massive quantum bosonic field interacting with a classical external source composed of charged particles. From $\hat{H}_1$, we can derive the Lagrangian for the classical field corresponding to the quantum bosonic field. On the other hand, $\hat{H}_2$ describes a massive fermionic field interacting with a classical external field. From $\hat{H}_2$, we can derive the Lagrangian for classical a particle moving in the external field.
In this section, we will first derive the classical limit of the quantum field theory described by the Hamiltonian $\hat{H}_1$.
Specifically, our goal is to determine the quantum states that correspond to classical fields and derive the Lagrangian of the classical fields from the quantum field theory described by the Hamiltonian $\hat{H}_1$ in \eqref{spllag}.

Let us recall that the free vacuum state $\left|\Omega\right>$ corresponds to the classical field $\phi_{\text{class}}(\boldsymbol{x},t)=0$ and $\pi_{\text{class}}(\boldsymbol{x},t)=0$. The wave function of the vacuum state is given by\cite{Weinberg2005}
\begin{equation}
	\begin{split}	
		\left<\phi|\Omega\right>=\mathcal{N}\exp\left\{-\frac{1}{2}\int d^3xd^3y \;
		\mathcal{E}(\boldsymbol{x},\boldsymbol{y})\phi(\boldsymbol{x})\phi(\boldsymbol{y})
		\right\}
		\; ,
	\end{split}
\end{equation}
where $\mathcal{N}$ is the normalization coefficient, $\left|\phi\right>$ is the eigenstate of the operator $\hat\phi(\boldsymbol{x})$ with the eigenvalue $\phi(\boldsymbol{x})$, and $\mathcal{E}(\boldsymbol{x},\boldsymbol{y})$ is the kernel defined by
\begin{equation}
	\label{E}
	\mathcal{E}(\boldsymbol{x},\boldsymbol{y})
	=\textcolor{black}{\frac{1}{\hbar^5}}\int\frac{d^3p}{(2\pi)^3} e^{\frac{i}{\textcolor{black}{\hbar}}\boldsymbol{p}\cdot(\boldsymbol{x}-\boldsymbol{y})} E_{\boldsymbol{p}}
	\; .
\end{equation}

The vacuum state wave function appears as a Gaussian wave packet with the center located at $\phi(\boldsymbol{x})=0$. We can generalize this wave function by shifting the center of the wave packet to $\phi(\boldsymbol{x})=\phi_{\text{class}}(\boldsymbol{x})$. Based on this generalization, we can write the general Gaussian wave ansatz as
\begin{equation}\label{gauwave}
	\begin{split}	
		\left<\phi|\varphi\right>=\mathcal{N}\exp\left\{-\frac{1}{2}\int d^3xd^3y \;
		\mathcal{E}(\boldsymbol{x},\boldsymbol{y})[\phi(\boldsymbol{x})-f(\boldsymbol{x})][\phi(\boldsymbol{y})-f(\boldsymbol{y})]
		\right\}
		\; ,
	\end{split}
\end{equation}
where $f(\boldsymbol{x})$ is an arbitrary function and $f(\boldsymbol{x})$ represents the expected center location for a non-vacuum state wave packet.

A quantum state that corresponds to a classical field will maintain its classical field correspondence during its time evolution. It will not evolve into a quantum state that lacks classical field correspondence. Therefore, the wave function should maintain the Gaussian ansatz during its evolution as follows:
\begin{equation}
	\label{assumption}
	\begin{split}	
		\left<\phi|\varphi(t)\right>
		=\mathcal{N}(t)\exp\left\{-\frac{1}{2}\int d^3xd^3y \;
		\mathcal{E}(\boldsymbol{x},\boldsymbol{y})[\phi(\boldsymbol{x})-f(\boldsymbol{x},t)][\phi(\boldsymbol{y})-f(\boldsymbol{y},t)]\right\}
		\; ,
	\end{split}
\end{equation}
where $\left|\varphi(t)\right>$ is a time-dependent state satified the Schrodinger equation $i\frac{\partial}{\partial t}\left|\varphi(t)\right>=\hat{H}\left|\varphi(t)\right>$, and $\hat{H}_1$ is the Hamiltonian in \eqref{spllag}.

It is worth noting that all states or wave functions are governed by the Schrodinger equation, which allows us to determine the normalization coefficient $\mathcal{N}(t)$ and the center function $f(\boldsymbol{x},t)$. The consistency between the Schrodinger equation and the general Gaussian wave function will validate our ansatz. The detailed calculations are presented in Appendix A, and here we provide the final results for the normalization coefficient $\mathcal{N}(t)$ and the center function $f(\boldsymbol{x},t)$ (see \eqref{Fnew}):
	\begin{equation}
	\label{Fnew1}	
	\begin{split}
		\mathcal{N}(t)=\mathcal{N}_0\exp\Bigg\{\frac{1}{2}&\int d^3xd^3y \;
		\mathcal{E}(\boldsymbol{x},\boldsymbol{y}) f(\boldsymbol{x},t)f(\boldsymbol{y},t)
		\\
		&+i\frac{\hbar}{2}\int_{t_0}^{t}d\tau\int d^3xd^3y \;
		\mathcal{E}_2(\boldsymbol{x},\boldsymbol{y} )
		f(\boldsymbol{x},\tau)f(\boldsymbol{y},\tau) \Bigg\}
		\; ,
	\end{split}
\end{equation}
where  $\mathcal{N}_0$ is the normalization constant, and $\mathcal{E}_{-1}(\boldsymbol{x},\boldsymbol{y})$ is a newly defined kernel function given by 
$	\mathcal{E}_{-1}(\boldsymbol{x},\boldsymbol{y})
\equiv \frac{1}{\hbar}\int\frac{d^3p}{(2\pi)^3}\frac{1}{E_{\boldsymbol{p}}} e^{\frac{i}{\hbar}\boldsymbol{p}\cdot(\boldsymbol{x}-\boldsymbol{y})}
$.
Furthermore, we can explicitly express the real and imaginary parts of $f(\boldsymbol{x},t)$ as
\begin{equation}
	\label{ffff}
	f(\boldsymbol{x},t)=f_1(\boldsymbol{x},t)+\frac{i}{\textcolor{black}{\hbar}}\int d^3y\;\mathcal{E}_{-1}(\boldsymbol{x},\boldsymbol{y} )f_2(\boldsymbol{y},t)
	\; ,
\end{equation}
where $f_1(\boldsymbol{x},t)$ and $f_2(\boldsymbol{x},t)$ are real functions that satisfy the following equations (see \eqref{f2}):
\begin{equation}
	\label{f}
	\begin{split}	
		&\dot f_1(\boldsymbol{x},t)=f_2(\boldsymbol{x},t) \; ,\\
		&\dot f_2(\boldsymbol{x},t)=c^2\nabla^2  f_1(\boldsymbol{x},t)
		-\left(\frac{m_0c^2}{\textcolor{black}{\hbar}}\right)^2f_1(\boldsymbol{x},t)-\hbar^{\frac{1}{2}}c^{\frac{3}{2}}g j(\boldsymbol{x},t)
		\; .
	\end{split}
\end{equation}

In order to gain a clearer understanding of the physical meaning of $f_1(\boldsymbol{x},t)$ and $f_2(\boldsymbol{x},t)$, we substitute \eqref{ffff} back into \eqref{assumption}, yielding
\begin{equation}
	\label{fff}
	\begin{split}	
		\left<\phi|\varphi(t)\right>
		\sim
		&\exp\left\{
		-\frac{1}{2}\int d^3xd^3y \;
		\mathcal{E}(\boldsymbol{x},\boldsymbol{y})
		\left[\phi(\boldsymbol{x})-f_1(\boldsymbol{x},t)\right]
		\left[\phi(\boldsymbol{y})-f_1(\boldsymbol{y},t)\right]\right\}\\
		&\times \exp\left\{\frac{i}{\hbar}\int d^3x\; f_2(\boldsymbol{x},t)\phi(\boldsymbol{x})
		\right\}
		\;  .
	\end{split}
\end{equation}
Since $f_1(\boldsymbol{x},t)$ and $f_2(\boldsymbol{x},t)$ (as well as $\mathcal{E}(\boldsymbol{x},\boldsymbol{y})$ and $\phi(\boldsymbol{x})$) are real, the first line in \eqref{fff} indicates that the wave function $\left<\phi|\varphi(t)\right>$ represents a Gaussian wave packet centered around $f_1(\boldsymbol{x},t)$. In quantum mechanics, usual wave packets include an additional term $\propto e^{i\boldsymbol{p}\cdot\boldsymbol{x}}$, which corresponds to the second line in \eqref{fff}. This correspondence suggests that we interpret $f_2(\boldsymbol{x},t)$ in \eqref{fff} as the conjugate momentum $\pi$. Furthermore, we will prove that the probability distribution of the wave function in the $\phi$-representation $|\left<\phi|\varphi(t)\right>|^2$ has a peak at $\phi(\boldsymbol{x})=f_1(\boldsymbol{x},t)$, while the probability distribution of the wave function in the $\pi$-representation $|\left<\pi|\varphi(t)\right>|^2$ has a peak at $\pi(\boldsymbol{x})=f_2(\boldsymbol{x},t)$. 
Therefore, we can deduce that quantum state $\left|\varphi(t)\right>$ corresponds to the classical fields with $\phi_{\text{class}}(\boldsymbol{x},t)\equiv f_1(\boldsymbol{x},t)$ and $\pi_{\text{class}}(\boldsymbol{x},t)\equiv f_2(\boldsymbol{x},t)$. By substituting $f_1(\boldsymbol{x},t)$ and $f_2(\boldsymbol{x},t)$ with $\phi_{\text{class}}(\boldsymbol{x},t)$ and $\pi_{\text{class}}(\boldsymbol{x},t)$ in \eqref{f}, we obtain the equations of motion for the classical fields
\begin{equation}
	\begin{split}	
		\label{cl}
		&\dot \phi_{\text{class}}(\boldsymbol{x},t)=\pi_{\text{class}}(\boldsymbol{x},t)\;, \\
		&\dot \pi_{\text{class}}(\boldsymbol{x},t)=\textcolor{black}{c^2}\nabla^2  \phi_{\text{class}}(\boldsymbol{x},t)-\left(\frac{m_0\textcolor{black}{c^2}}{\textcolor{black}{\hbar}}\right)^2 \phi_{\text{class}}(\boldsymbol{x},t)-\textcolor{black}{\hbar^{\frac{1}{2}}c^{\frac{3}{2}}}g j(\boldsymbol{x},t)\; .
	\end{split}
\end{equation}
The equations \eqref{cl} can be interpreted as the Hamiltonian canonical equations for classical fields, and the corresponding Hamiltonian is given by
\begin{equation}
	\label{HHHH}
	H_{\text{class}}=\int d^3x\left[\frac{1}{2}\pi_{\text{class}}^2+\frac{\textcolor{black}{c^2}}{2}\left(\nabla\phi_{\text{class}}\right)^2+\frac{1}{2}\left(\frac{m_0\textcolor{black}{c^2}}{\textcolor{black}{\hbar}}\right)^2\phi_{\text{class}}^2+\textcolor{black}{\hbar^{\frac{1}{2}}c^{\frac{3}{2}}}gj\phi_{\text{class}} \right]\; .
\end{equation}
The Lagrangian corresponding to \eqref{HHHH} is
\begin{equation}\label{L1}
	\begin{split}
		\mathcal{L}_1=\frac{1}{2}\partial_{\mu}\phi_{\text{class}}\partial^{\mu}\phi_{\text{class}}-\frac{1}{2}\left(\frac{m_0c^2}{\hbar}\right)^2\phi_{\text{class}}^2-\hbar^{\frac{1}{2}}c^{\frac{3}{2}}gj\phi_{\text{class}}\; .
	\end{split}
\end{equation}

Let's analyze the properties of the wave function in more detail. By substituting $f_1(\boldsymbol{x},t)$ and $f_2(\boldsymbol{x},t)$ with $\phi_{\text{class}}(\boldsymbol{x},t)$ and $\pi_{\textcolor{black}{\text{class}}}(\boldsymbol{x},t)$, we can rewrite the time-dependent wave function \eqref{assumption} as
\begin{equation}
	\begin{split}
		\label{phi state}	
		\left<\phi|\varphi(t)\right>
		=\mathcal{N'}(t)
		&\exp\left\{
		-\frac{1}{2}\int d^3xd^3y \;
		\mathcal{E}(\boldsymbol{x},\boldsymbol{y})
		\left[\phi(\boldsymbol{x})-\phi_{\textcolor{black}{\text{class}}}(\boldsymbol{x},t)\right]
		\left[\phi(\boldsymbol{y})-\phi_{\textcolor{black}{\text{class}}}(\boldsymbol{y},t)\right]\right\}\\
		&\times \exp\left\{\frac{i}{\textcolor{black}{\hbar}}\int d^3x\;\pi_{\textcolor{black}{\text{class}}}(\boldsymbol{x},t)\phi(\boldsymbol{x})
		\right\}
		\; ,
	\end{split}
\end{equation}
where
	\begin{equation}	\label{Fnew2}
	\begin{split}
		\mathcal{N'}(t)
		=\mathcal{N}_0
		\exp\Bigg\{		
		\frac{1}{2}&\int d^3xd^3y \;
		\mathcal{E}(\boldsymbol{x},\boldsymbol{y})
		\phi_{\text{class}}(\boldsymbol{x},t)\phi_{\text{class}}(\boldsymbol{y},t)\\
		&		
		-\int_{t_0}^{t}d\tau
		\int d^3xd^3y \;
		\mathcal{E}(\boldsymbol{x},\boldsymbol{y})
		\phi_{\text{class}}(\boldsymbol{x},t)
		\pi_{\text{class}}(\boldsymbol{y},t)
		\Bigg\}\\
		\times\exp\Bigg\{&
		i\frac{\hbar}{2}\int_{t_0}^{t}d\tau
		\int d^3xd^3y \;
		\mathcal{E}_2(\boldsymbol{x},\boldsymbol{y})
		\phi_{\text{class}}(\boldsymbol{x},t)\phi_{\text{class}}(\boldsymbol{y},t)\\
		&~~~~~~~~-i\frac{1}{2\hbar}\int_{t_0}^{t}d\tau
		\int d^3x\;
		\pi_{\text{class}}(\boldsymbol{x},t)\pi_{\text{class}}(\boldsymbol{x},t)
		\Bigg\}
		\; .
	\end{split}
\end{equation}
According to \eqref{cl}, we can see that the first $\exp\{\cdots\}$ term in \eqref{Fnew2} is a constant. Therefore, \eqref{Fnew2} can be further written as
\begin{equation}	\label{Fnew3}
	\begin{split}
		\mathcal{N'}(t)
		=\mathcal{N}'_0
		\exp\Bigg\{
		i\frac{\hbar}{2}\int_{t_0}^{t}d\tau
		&\int d^3xd^3y \;
		\mathcal{E}_2(\boldsymbol{x},\boldsymbol{y})
		\phi_{\text{class}}(\boldsymbol{x},t)\phi_{\text{class}}(\boldsymbol{y},t)\\
		&-i\frac{1}{2\hbar}\int_{t_0}^{t}d\tau
		\int d^3x\;
		\pi_{\text{class}}(\boldsymbol{x},t)\pi_{\text{class}}(\boldsymbol{x},t)
		\Bigg\}
		\; ,
	\end{split}
\end{equation}
where $\mathcal{N}'_0$ is independent of time.
Substituting \eqref{Fnew3} into \eqref{phi state}, we can obtain the probability distribution function of the variable $\phi$ as
\begin{equation}
	\begin{split}
		\label{prob}	
		|\left<\phi|\varphi(t)\right>|^2
		=|\mathcal{N}'_0|^2
		\exp\left\{
		-\int d^3xd^3y \;
		\mathcal{E}(\boldsymbol{x},\boldsymbol{y})
		\left[\phi(\boldsymbol{x})-\phi_{\text{class}}(\boldsymbol{x},t)\right]
		\left[\phi(\boldsymbol{y})-\phi_{\text{class}}(\boldsymbol{y},t)\right]\right\}
		\;  .
	\end{split}
\end{equation}

To confirm that the wave packet is localized around $\phi_{\text{class}}$, we carefully examine the probability distribution \eqref{prob}. For convenience, we define the functional $F[\phi]$ as
\begin{equation}
	\begin{split}
		F[\phi]\equiv
		\int d^3xd^3y \;
		\mathcal{E}(\boldsymbol{x},\boldsymbol{y})
		\Delta \phi(\boldsymbol{x}) \Delta \phi(\boldsymbol{y})
		\; ,
	\end{split}
\end{equation}
where $\Delta \phi(\boldsymbol{x})=\phi(\boldsymbol{x})-\phi_{\textcolor{black}{\text{class}}}(\boldsymbol{x},t)$
and $\Delta \phi(\boldsymbol{y})=\phi(\boldsymbol{y})-\phi_{\textcolor{black}{\text{class}}}(\boldsymbol{y},t)$. We consider the Fourier transformation for $\phi(\boldsymbol{x},t)$
\begin{equation}
	\begin{split}
		\Delta \phi(\boldsymbol{x})=\int d^3k\; \phi(\boldsymbol{k})e^{\frac{i}{\textcolor{black}{\hbar}}\boldsymbol{k}\cdot\boldsymbol{x}}
		\; .
	\end{split}
\end{equation}
Then, $F[\phi]$ can be expressed in terms of $\phi(\boldsymbol{k})$
\begin{equation}
	\begin{split}
		F[\phi]
		=\textcolor{black}{\hbar c}\int d^3k \; 
		\sqrt{\boldsymbol{k} ^2+m^2\textcolor{black}{c^2}}
		|\phi(\boldsymbol{k})|^2
		\geq 0
		\; .
	\end{split}
\end{equation}
The non-negative property of $F[\phi]$ indicates that $F[\phi(\boldsymbol{x})]$ reaches its minimum when $\Delta \phi(\boldsymbol{x})=\phi(\boldsymbol{x})-\phi_{\text{class}}(\boldsymbol{x},t)=0$. This means that $|\left<\phi|\varphi(t)\right>|^2\sim e^{-F[\phi]}$ decreases as $\phi$ moves away from $\phi_{\text{class}}$. Therefore, we can conclude that the wave function represents a wave packet centered around the classical field $\phi_{\text{class}}(\boldsymbol{x},t)$.

The remaining question is whether the wave function is centered around $\pi_{\textcolor{black}{\text{class}}}(\boldsymbol{x},t)$ in the $\pi$-representation, as the expression for \eqref{phi state} does not provide information about the centering around $\pi_{\textcolor{black}{\text{class}}}(\boldsymbol{x},t)$. Therefore, we need to examine the wave function in the $\pi$-representation. The full calculation is presented in Appendix \eqref{sec:AppB}, and we present the resulting wave function here
\begin{equation}
	\begin{split}	
		\label{pi*}
		\left<\pi|\varphi(t)\right>
		\sim
		\exp\left\{-\frac{1}{2\textcolor{black}{\hbar^2}}\int d^3x d^3y \; \mathcal{E}_{-1}(\boldsymbol{x},\boldsymbol{y})
		[\pi(\boldsymbol{x})-\pi_{\textcolor{black}{\text{class}}}(\boldsymbol{x},t)][\pi(\boldsymbol{y})-\pi_{\textcolor{black}{\text{class}}}(\boldsymbol{y},t)] \right\}
		\; .
	\end{split}
\end{equation}
Clearly, $\left<\pi|\varphi(t)\right>$ is a wave packet centered around the classical field conjugate $\pi_{\textcolor{black}{\text{class}}}(\boldsymbol{x},t)$. Therefore, the quantum state $\left|\varphi(t)\right>$ is indeed a Gaussian wave function that corresponds to the classical field composed of $\phi_{\textcolor{black}{\text{class}}}(\boldsymbol{x},t)$ and $\pi_{\textcolor{black}{\text{class}}}(\boldsymbol{x},t)$. The equation for the center function coincides with the equation of motion for classical fields.

In fact, the wave packet \eqref{gauwave} corresponds to a well-known quantum state called the coherent state. The field operators $\hat\phi$ and $\hat\pi$ can be expanded in terms of the annihilation operator, and the annihilation operator can be expressed in terms of the field operators as follows
\begin{equation}
	\begin{split}
		\hat a_{\boldsymbol{p}}=\int d^3x\;
		e^{-i\boldsymbol{p}\cdot\boldsymbol{x}}
		\left[\sqrt{\frac{E_{\boldsymbol{p}}}{2}}\hat\phi(\boldsymbol{x})
		+\frac{i}{\sqrt{2E_{\boldsymbol{p}}}}\hat\pi(\boldsymbol{x})\right]
		\; .
	\end{split}
\end{equation}
Here, we ignore the factors of $\hbar$ and $c$ as they are not crucial for the discussion of coherent states.
Using Equation \eqref{hatpi} and the definition of $\mathcal{E}(\boldsymbol{x},\boldsymbol{y})$,  we have
\begin{equation}
	\begin{split}
		\label{ap}
		\hat a_{\boldsymbol{p}}\left<\phi|\varphi\right>
		&=\int d^3x\;
		e^{-i\boldsymbol{p}\cdot\boldsymbol{x}}
		\left[\sqrt{\frac{E_{\boldsymbol{p}}}{2}}\phi(\boldsymbol{x})
		-\frac{1}{\sqrt{2E_{\boldsymbol{p}}}}\int d^3y 
		\mathcal{E}(\textbf x,\textbf y)[\phi(\textbf y)-f(\textbf y)]\right]
		\left<\phi|\varphi\right>\\
		&=\left[\frac{1}{\sqrt{2E_{\boldsymbol{p}}}}
		\int d^3x
		e^{-i\boldsymbol{p}\cdot\boldsymbol{x}}
		\int d^3y 
		\mathcal{E}(\textbf x,\textbf y)f(\textbf y)\right]
		\left<\phi|\varphi\right>
		\; .
	\end{split}
\end{equation}
This equation indicates that the wave packet \eqref{gauwave} is an eigenstate of the annihilation operator $\hat a_{\boldsymbol{p}}$, which means that the quantum state we are considering here is a coherent state.

Let's further discuss the structure of the quantum state \eqref{phi state}, which is actually a coherent state in the representation expanded by the eigenstates of the quantum field operators.

The field operator $\hat\phi$ at different points are independent of each other and satisfy the commutation relations $\left[\hat\phi(x),\hat\phi(y)\right]=0$. As a result, the eigenstate of $\hat\phi$ can be written in a direct product formulation as follows
$$|\phi\rangle=
|\phi(\boldsymbol{x_1})\rangle |\phi(\boldsymbol{x_2})\rangle
|\phi(\boldsymbol{x_3})\rangle  \cdots
=\prod_{\boldsymbol{x}}  |\phi(\boldsymbol{x})\rangle\; .$$
Here, the quantum state $|\phi(\boldsymbol{x})\rangle$ is the eigenstate of $\hat\phi(\boldsymbol{x})$ but in a Hilbert space constructed only for the point $\boldsymbol{x}$. It is evident that the quantum state $|\phi\rangle$ is not an entangled state. However, the state given by \eqref{phi state} is an entangled state, which can be seen from the following form
\begin{equation}
	\begin{split}
		\label{ephi state}	
		|\varphi(t)\rangle
		=&\mathcal{N'}(t)
		\left(\prod_{\boldsymbol{x}}\int d\phi(\boldsymbol{x})\right)
		\exp\left\{\frac{i}{\hbar}\int d^3x \; \pi_{\text{class}}(\boldsymbol{x},t)\phi(\boldsymbol{x})
		\right\}\\
		&\times \exp\left\{
		-\frac{1}{2}\int d^3xd^3y \;
		\mathcal{E}(\boldsymbol{x},\boldsymbol{y})
		\left[\phi(\boldsymbol{x})-\phi_{\text{class}}(\boldsymbol{x},t)\right]
		\left[\phi(\boldsymbol{y})-\phi_{\text{class}}(\boldsymbol{y},t)\right]\right\}
		\prod_{\boldsymbol{x}}  |\phi(\boldsymbol{x})\rangle
		\;  .
	\end{split}
\end{equation}
Due to the presence of the term $\int d^3x d^3y
\mathcal{E}(\boldsymbol{x},\boldsymbol{y})
\left[\phi(\boldsymbol{x})-\phi_{\text{class}}(\boldsymbol{x},t)\right]
\left[\phi(\boldsymbol{y})-\phi_{\text{class}}(\boldsymbol{y},t)\right]$ in \eqref{ephi state}, $|\varphi(t)\rangle$ must be expressed as a summation and cannot be written solely as a direct product. This might seem to contradict the classical theory, where a classical field is a local field with a single value at each point, and the values at different points are not entangled with each other. However, the behavior of $\mathcal{E}(\boldsymbol{x},\boldsymbol{y})\propto e^{-m|\boldsymbol{x}-\boldsymbol{y}|}\to0$ as $|\boldsymbol{x}-\boldsymbol{y}|\to\infty$ implies that the entanglement between points with large spatial separations becomes weak. This means that measurements of the state \eqref{ephi state} at different points will not interfere with each other from a macroscopic perspective. Therefore, it is reasonable to interpret the wave function \eqref{phi state} as the quantum correspondence of the classical field.

The introduction of the quantum superposition principle is often illustrated through the double-slit interference phenomenon in optics in textbooks. Initially, the classical double-slit experiment is presented, where light is described by classical fields. Then, by reducing the intensity of the light source to the point where only one dot appears on the screen at a time, the classical experiment transitions into a quantum one. Over time, with the accumulation of dots, the interference fringes of classical wave optics emerge. This approach introduces the idea of the quantum superposition principle based on the classical superposition principle of classical fields. However, this could potentially mislead us into thinking that classical superposition is an amplified version of quantum superposition.
To highlight the difference between quantum and classical superposition, let's consider a classical field $\phi_{1+2}$ as the superposition of two classical fields $\phi_1$ and $\phi_2$, i.e., $\phi_{1+2}=\phi_1+\phi_2$. We can then express the wave function of $\phi_{1+2}$ in the $\phi$-representation as follows
\begin{equation}\label{superposition}
	\begin{split}
		&\left<\phi|\varphi_{1+2}(t)\right>
		=\mathcal{N'}(t) \exp\left\{\frac{i}{\textcolor{black}{\hbar}}\int d^3x\;[\pi_1(\boldsymbol{x},t)+\pi_2(\boldsymbol{x},t)]\phi(\boldsymbol{x})
		\right\}\\
		&\times\exp\left\{
		-\frac{1}{2}\int d^3xd^3y \;
		\mathcal{E}(\boldsymbol{x},\boldsymbol{y})
		\left[\phi(\boldsymbol{x})-\phi_1(\boldsymbol{x},t)-\phi_2(\boldsymbol{x},t)\right]
		\left[\phi(\boldsymbol{y})-\phi_1(\boldsymbol{y},t)-\phi_2(\boldsymbol{y},t)\right]\right\}
		\; .
	\end{split}
\end{equation}
The wave functions of $\phi_1$ and $\phi_2$ are denoted as $\left<\phi|\varphi_{1}(t)\right>$ and $\left<\phi|\varphi_{2}(t)\right>$, respectively, which can be easily obtained using \eqref{phi state}. From the explicit expression of the wave function $\left<\phi|\varphi_{1+2}(t)\right>$ in the $\phi$-representation, it is apparent that the wave function for the superposition of two classical fields is not the superposition of the corresponding wave functions:
\begin{equation}\label{dj}
	\begin{split}
		\left<\phi|\varphi_{1+2}(t)\right>
		\neq
		k\Big[\left<\phi|\varphi_{1}(t)\right>+\left<\phi|\varphi_{2}(t)\right>\Big]
		\;  ,
	\end{split}
\end{equation}
where $k$ represents the normalization coefficient. 
In fact, for any quantum superposition coefficients $c_1$ and $c_2$, there exists the inequality $\left<\phi|\varphi_{1+2}(t)\right> \neq c_1\left<\phi|\varphi_{1}(t)\right>+c_2\left<\phi|\varphi_{2}(t)\right>$. We know that classical superposition can produce interference patterns on the screen consistent with experiments, a process represented as $\phi_{1+2}=\phi_1+\phi_2$. The left side of Eq. \eqref{dj} can be interpreted as the quantum representation of classical superposition. Computing the average value of the field operator $\hat\phi$ for the quantum state $|\varphi_{1+2}(t)\rangle$ yields interference patterns consistent with classical superposition, i.e., $\langle \varphi_{1+2}(t)|\hat\phi|\varphi_{1+2}(t)\rangle =\phi_{1+2}=\phi_1+\phi_2$, with very small uncertainty about the operator $\hat\phi$. 
On the other hand, the right hand side of Eq. \eqref{dj} corresponds to the quantum state after  quantum superposition, $k\left(|\varphi_{1}(t)\rangle+|\varphi_{2}(t)\rangle\right)$. Computing the average value of the field operator $\hat\phi$ for this quantum state, the result is obviously not equal to $\phi_{1+2}=\phi_1+\phi_2$, and the uncertainty in this case is relatively large. In fact, it represents a physical system that cannot be described by classical fields.
Therefore, we cannot interpret classical superposition as a naive macroscopic version of quantum superposition.


\section{Reduced density matrices of the quantum states corresponding to classical fields}
	\label{red}
	
	In addition to the various properties discussed earlier regarding the quantum state \eqref{phi state}, this section will introduce a more significant property.
One major characteristic of classical fields is that there is a definite value at each point in space, allowing us to explicitly discuss the distribution of the field in a particular spatial region without concerning ourselves with fields outside that region.
However, the quantum state \eqref{phi state}, corresponding to classical fields, exhibits spatial entanglement, lacking a natural notion of locality as classical fields do. For a region, according to the Ref. \cite{Tu2024}, we can use the reduced density matrices to describe the information of the quantum state on that region.
This leads to a crucial question:
if two classical fields are identical in a certain region, are the reduced density matrices of their corresponding quantum states also identical in that region?
For example, dividing the entire space into two regions, denoted as $A$ and $a$, if a classical field in region $A$ is in the classical vacuum state (i.e., $\phi_{\text{class}}(\boldsymbol{x}_A,t)= \pi_{\text{class}}(\boldsymbol{x}_A,t)=0$ for $\boldsymbol{x}_A\in A$), then is the quantum state $|\varphi\rangle$ corresponding to this classical field equivalent to the vacuum state $|\Omega\rangle$  in region $A$ (i.e., $\text{tr}_a\left(|\varphi\rangle \langle\varphi|\right) = \text{tr}_a\left(|\Omega\rangle \langle\Omega|\right)$)?

	To address the above question, we need to compute the reduced density matrix explicitly.
	Based on the quantum state \eqref{phi state} corresponding to the classical field $\phi_{\text{class}}(\boldsymbol{x},t)$ and $\pi_{\text{class}}(\boldsymbol{x},t)$, along with the normalization coefficient \eqref{Fnew3}, we can write the density matrix in the entire space as
	\begin{equation}
		\begin{split}
			&\rho(\phi,\phi'; t)	\\
			&=\left<\phi|\varphi(t)\right> 	\left<\varphi(t)|\phi'\right>\\
			&	=|\mathcal{N}'_0|^2
			\exp\left\{
			-\frac{1}{2}\int d^3xd^3y \;
			\mathcal{E}(\boldsymbol{x},\boldsymbol{y})
			\left[\phi(\boldsymbol{x})-\phi_{\text{class}}(\boldsymbol{x},t)\right]
			\left[\phi(\boldsymbol{y})-\phi_{\text{class}}(\boldsymbol{y},t)\right]\right\}\\
			&~~~~~~~~~~~\times \exp\left\{\frac{i}{\hbar}\int d^3x \; \pi_{\text{class}}(\boldsymbol{x},t)
			\Big[\phi(\boldsymbol{x})-\phi'(\boldsymbol{x})\Big]
			\right\}
			\\
			&~~~~~~~~~~~	\times	\exp\left\{
			-\frac{1}{2}\int d^3xd^3y \;
			\mathcal{E}(\boldsymbol{x},\boldsymbol{y})
			\left[\phi'(\boldsymbol{x})-\phi_{\text{class}}(\boldsymbol{x},t)\right]
			\left[\phi'(\boldsymbol{y})-\phi_{\text{class}}(\boldsymbol{y},t)\right]\right\}
			\; .
		\end{split}
	\end{equation}
Dividing the space into two regions, denoted as region $A$ and region $a$, and tracing out region $a$, we can obtain the reduced density matrix in region $A$ as
\begin{equation}\label{den}
	\begin{split}
	\int\mathcal{D}\phi_a\;	\rho(\phi,&\phi';t)
		\bigg|_{\phi(\boldsymbol{x}_a)=\phi'(\boldsymbol{x}_a),\;\boldsymbol{x}_a\in a}	\\
				=|\mathcal{N}'_0|^2
		\int\mathcal{D}&\phi_a\;
		\exp\left\{-\int_a d^3x  \int_a d^3y \;
		\mathcal{E}(\boldsymbol{x}_a,\boldsymbol{y}_a)
		\left[\phi(\boldsymbol{x}_a)-\phi_{\text{class}}(\boldsymbol{x}_a,t)\right]
		\left[\phi(\boldsymbol{y}_a)-\phi_{\text{class}}(\boldsymbol{y}_a,t)\right]\right\}
		\\&
		\times\exp\left\{
		-\int_a d^3x  \int_A d^3y \;
		\mathcal{E}(\boldsymbol{x}_a ,\boldsymbol{y}_A)
		\left[\phi(\boldsymbol{x}_a )-\phi_{\text{class}}(\boldsymbol{x}_a ,t)\right]
		\left[\phi(\boldsymbol{y}_A)-\phi_{\text{class}}(\boldsymbol{y}_A,t)\right]\right\}
		\\&
		\times\exp\left\{
		-\frac{1}{2}\int_A d^3x\int_A d^3y \;
		\mathcal{E}(\boldsymbol{x}_A,\boldsymbol{y}_A)
		\left[\phi(\boldsymbol{x}_A)-\phi_{\text{class}}(\boldsymbol{x}_A,t)\right]
		\left[\phi(\boldsymbol{y}_A)-\phi_{\text{class}}(\boldsymbol{y}_A,t)\right]\right\}
		\\&
		\times\exp\left\{
		-\int_a d^3x  \int_A d^3y \;
		\mathcal{E}(\boldsymbol{x}_a,\boldsymbol{y}_A)
		\left[\phi(\boldsymbol{x}_a)-\phi_{\text{class}}(\boldsymbol{x}_a,t)\right]
		\left[\phi'(\boldsymbol{y}_A)-\phi_{\text{class}}(\boldsymbol{y}_A,t)\right]\right\}
		\\&
		\times\exp\left\{
		-\frac{1}{2}\int_A d^3x\int_A d^3y \;
		\mathcal{E}(\boldsymbol{x}_A,\boldsymbol{y}_A)
		\left[\phi'(\boldsymbol{x}_A)-\phi_{\text{class}}(\boldsymbol{x}_A,t)\right]
		\left[\phi'(\boldsymbol{y}_A)-\phi_{\text{class}}(\boldsymbol{y}_A,t)\right]\right\}
		\\
		&\times \exp\left\{\frac{i}{\hbar}\int_A d^3x\;\pi_{\text{class}}(\boldsymbol{x}_A,t)
		\Big[\phi(\boldsymbol{x}_A)-\phi'(\boldsymbol{x}_A)\Big]
		\right\}
		\;,
	\end{split}
\end{equation}
where $\int\mathcal{D}\phi_a$ represents the integration over only the fields $\phi$ in region $a$. We can further perform a variable substitution $\varphi(\boldsymbol{x}_a)=\phi(\boldsymbol{x}_a)-\phi_{\text{class}}(\boldsymbol{x}_a,t)$ to simplify \eqref{den} into the following form:
\begin{equation}\label{den1}
	\begin{split}
		\int\mathcal{D}\phi_a\;&\rho(\phi,\phi';t)
		\bigg|_{\phi(\boldsymbol{x}_a)=\phi'(\boldsymbol{x}_a),\;\boldsymbol{x}_a\in a}	\\
			=|\mathcal{N}'_0|^2
			&\exp\left\{\frac{i}{\hbar}\int_A d^3x\;\pi_{\text{class}}(\boldsymbol{x}_A,t)
		\Big[\phi(\boldsymbol{x}_A)-\phi'(\boldsymbol{x}_A)\Big]
		\right\}
		\\&
		\times\exp\left\{
		-\frac{1}{2}\int_A d^3x\int_A d^3y \;
		\mathcal{E}(\boldsymbol{x}_A,\boldsymbol{y}_A)
		\left[\phi(\boldsymbol{x}_A)-\phi_{\text{class}}(\boldsymbol{x}_A,t)\right]
		\left[\phi(\boldsymbol{y}_A)-\phi_{\text{class}}(\boldsymbol{y}_A,t)\right]\right\}
		\\&
		\times\exp\left\{
		-\frac{1}{2}\int_A d^3x\int_A d^3y \;
		\mathcal{E}(\boldsymbol{x}_A,\boldsymbol{y}_A)
		\left[\phi'(\boldsymbol{x}_A)-\phi_{\text{class}}(\boldsymbol{x}_A,t)\right]
		\left[\phi'(\boldsymbol{y}_A)-\phi_{\text{class}}(\boldsymbol{y}_A,t)\right]\right\}
		\\		
		\times\int\mathcal{D}&\varphi_a\;
		\exp\left\{-\int_a d^3x  \int_a d^3y\;
		\mathcal{E}(\boldsymbol{x}_a,\boldsymbol{y}_a)
		\varphi(\boldsymbol{x}_a)\varphi(\boldsymbol{y}_a)\right\}
		\\&
		\times\exp\left\{
		-\int_a d^3x  \int_A d^3y \;
		\mathcal{E}(\boldsymbol{x}_a ,\boldsymbol{y}_A)
		\varphi(\boldsymbol{x}_a)
		\left[\phi(\boldsymbol{y}_A)-\phi_{\text{class}}(\boldsymbol{y}_A,t)\right]\right\}
		\\&
		\times\exp\left\{
		-\int_a d^3x  \int_A d^3y \;
		\mathcal{E}(\boldsymbol{x}_a,\boldsymbol{y}_A)
		\varphi(\boldsymbol{x}_a)
		\left[\phi'(\boldsymbol{y}_A)-\phi_{\text{class}}(\boldsymbol{y}_A,t)\right]\right\}
		\;,
	\end{split}
\end{equation}
where $\boldsymbol{x}_A\in A$ and $\boldsymbol{x}_a\in a$.
Note that in \eqref{den1}, there are no appearances of $\phi_{\text{class}}(\boldsymbol{x}_a,t)$ and $\pi_{\text{class}}(\boldsymbol{x}_a,t)$. This indicates that the reduced density matrix of the quantum state \eqref{phi state} in region $A$ only depends on the classical fields $\phi_{\text{class}}(\boldsymbol{x}_A,t)$ and $\pi_{\text{class}}(\boldsymbol{x}_A,t)$ in region $A$. Therefore, if two classical fields are identical in a certain region, then the reduced density matrices of their corresponding quantum states are also identical in that region. This self-consistency demonstrates that the quantum state \eqref{phi state} indeed perfectly corresponds to the classical field, and spatial entanglement and non-locality of the quantum state do not disturb this correspondence. Additionally, we obtain a strong corollary: knowing the classical field in a local region is sufficient to immediately write out the corresponding reduced density matrix \eqref{den1} for that region.
Thus, classical fields and quantum descriptions not only have global correspondences but also have local correspondences, and all correspondences are self-consistent.

Although the spatial entanglement of quantum states does not destroy the consistency of the correspondence between quantum and classical fields, it does lead to other counterintuitive conclusions.  Consider two classical field distributions at $t=0$, denoted as $\phi_{\text{class}}(\boldsymbol{x}),\pi_{\text{class}}(\boldsymbol{x})$ and $\phi'_{\text{class}}(\boldsymbol{x}),\pi'_{\text{class}}(\boldsymbol{x})$. Using \eqref{phi state}, we obtain their corresponding quantum states $|\varphi\rangle$ and $|\varphi'\rangle$, respectively.
Suppose these two classical fields are both in vacuum state in region $A$:
$$\phi_{\text{class}}(\boldsymbol{x}_A)= \pi_{\text{class}}(\boldsymbol{x}_A)=0
=\phi'_{\text{class}}(\boldsymbol{x}_A)= \pi'_{\text{class}}(\boldsymbol{x}_A)
,\;\boldsymbol{x}_A\in A\;.$$
This implies that $|\varphi\rangle$ and $|\varphi'\rangle$ in region $A$ are equivalent to the vacuum state $|\Omega\rangle$.
The problem we want to investigate is whether the superposition state $|\varphi_{+}\rangle=\frac{1}{\sqrt 2}\left(|\varphi\rangle+|\varphi'\rangle\right)$ remains equivalent to the vacuum state $|\Omega\rangle$ in region $A$.

According to \eqref{dj}, we know that quantum superposition is not equivalent to classical superposition. Therefore, although the classical fields superposed in region $A$ result in vacuum, it does not imply that the quantum state $|\varphi_{+}\rangle$ is equivalent to the vacuum state $|\Omega\rangle$ in region $A$.
To answer this question, we need to calculate the reduced density matrix of $|\varphi_{+}\rangle$ in region $A$:
\begin{equation}\label{dj1}
	\begin{split}
		&\int\mathcal{D}\phi_a\;
		\left<\phi|\varphi_+\right>	\left<\varphi_+|\phi'\right>		
		\bigg|_{\phi(\boldsymbol{x}_a)=\phi'(\boldsymbol{x}_a),\;\boldsymbol{x}_a\in a}	\\
		&=\frac{1}{2}\int\mathcal{D}\phi_a\;
		\left<\phi|\varphi\right>	\left<\varphi|\phi'\right>		
		\bigg|_{\phi(\boldsymbol{x}_a)=\phi'(\boldsymbol{x}_a),\;\boldsymbol{x}_a\in a}	
		+\frac{1}{2}\int\mathcal{D}\phi_a\;
		\left<\phi|\varphi'\right>	\left<\varphi'|\phi'\right>		
		\bigg|_{\phi(\boldsymbol{x}_a)=\phi'(\boldsymbol{x}_a),\;\boldsymbol{x}_a\in a}	
		\\&
		+\frac{1}{2}\int\mathcal{D}\phi_a\;
		\left<\phi|\varphi\right>	\left<\varphi'|\phi'\right>		
		\bigg|_{\phi(\boldsymbol{x}_a)=\phi'(\boldsymbol{x}_a),\;\boldsymbol{x}_a\in a}	
		+\frac{1}{2}\int\mathcal{D}\phi_a\;
		\left<\phi|\varphi'\right>	\left<\varphi|\phi'\right>		
		\bigg|_{\phi(\boldsymbol{x}_a)=\phi'(\boldsymbol{x}_a),\;\boldsymbol{x}_a\in a}	
		\; .
	\end{split}
\end{equation}
According to \eqref{den1}, the first line after the equality sign in \eqref{dj1} equals the reduced density matrix of the vacuum state in region $A$:
\begin{equation}\label{dj2}
	\begin{split}
		&\frac{1}{2}\int\mathcal{D}\phi_a\;
		\left<\phi|\varphi\right>	\left<\varphi|\phi'\right>		
		\bigg|_{\phi(\boldsymbol{x}_a)=\phi'(\boldsymbol{x}_a),\;\boldsymbol{x}_a\in a}	
		+\frac{1}{2}\int\mathcal{D}\phi_a\;
		\left<\phi|\varphi'\right>	\left<\varphi'|\phi'\right>		
		\bigg|_{\phi(\boldsymbol{x}_a)=\phi'(\boldsymbol{x}_a),\;\boldsymbol{x}_a\in a}	\\
		&=
		\int\mathcal{D}\phi_a\;
		\left<\phi|\Omega\right>	\left<\Omega|\phi'\right>		
		\bigg|_{\phi(\boldsymbol{x}_a)=\phi'(\boldsymbol{x}_a),\;\boldsymbol{x}_a\in a}	
		\; .
	\end{split}
\end{equation}
So, for \eqref{dj1} to equal the reduced density matrix of the vacuum in region $A$, the last line of \eqref{dj1} needs to be zero.

 Note that the two terms in the last line of \eqref{dj1} are each other's complex conjugates.
For specific classical field distributions, such as $ \pi_{\text{class}}(\boldsymbol{x})=\phi'_{\text{class}}(\boldsymbol{x})= \pi'_{\text{class}}(\boldsymbol{x})=0$, it can be proven from \eqref{phi state} that the phase of $\int\mathcal{D}\phi_a\;
\left<\phi|\varphi\right>	\left<\varphi'|\phi'\right>		
\bigg|_{\phi(\boldsymbol{x}_a)=\phi'(\boldsymbol{x}_a),\;\boldsymbol{x}_a\in a}	$ is independent of $\phi_A$ and $\phi'_A$. Hence, suitable normalization constant $\mathcal{N}'_0$ can be chosen to make $\int\mathcal{D}\phi_a\;
\left<\phi|\varphi\right>	\left<\varphi'|\phi'\right>		
\bigg|_{\phi(\boldsymbol{x}_a)=\phi'(\boldsymbol{x}_a),\;\boldsymbol{x}_a\in a}	$ purely imaginary, and in this case, \eqref{dj1} equals the reduced density matrix of the vacuum in region $A$.
However, for general classical fields, the phase of $\int\mathcal{D}\phi_a\;
\left<\phi|\varphi\right>	\left<\varphi'|\phi'\right>		
\bigg|_{\phi(\boldsymbol{x}_a)=\phi'(\boldsymbol{x}_a),\;\boldsymbol{x}_a\in a}	$ varies with $\phi_A$ and $\phi'_A$. Therefore, in most cases, \eqref{dj1} does not equal the reduced density matrix of the vacuum in region $A$.
In other words, even if $|\varphi\rangle$ and $|\varphi'\rangle$ are equivalent to the vacuum state $|\Omega\rangle$ in region $A$, their superposition state $|\varphi_{+}\rangle=\frac{1}{\sqrt 2}\left(|\varphi\rangle+|\varphi'\rangle\right)$ is no longer equivalent to the vacuum state $|\Omega\rangle$ in region $A$. 
In fact, through more careful calculation, it is easy to extend the above conclusion to:
when two quantum states with the same reduced density matrix in a certain region are superposed, the resulting superposition state no longer preserves the original reduced density matrix in that region due to the spatial entanglement.

\section{relativistic classical Yukawa theory}
\label{cy}

In this section, we derive the classical Lagrangian corresponding to the Hamiltonian $\hat{H}_2$ as shown in \eqref{spllag}. Classical theory does not involve fermionic fields; instead, fermions are treated as point particles with definite positions and momenta. Therefore, we only need to consider the Hilbert subspace consisting of one-particle states, where quantum field theory can be reduced to one-particle quantum mechanics. By employing wave packets peaking at a certain point in phase space, the concept of a point particle emerges in the limit $\hbar\to 0$, allowing us to obtain the classical limit of the one-particle quantum mechanics \cite{Cohn1972a,Kazandjian2006,Kazandjian2007,Klein2012}.

The equation of motion for a fermionic field in the Heisenberg picture can be derived from Hamiltonian $\hat{H}_2$ in \eqref{spllag}
\begin{equation}\label{eomope}	
	i\hbar\frac{\partial}{\partial t} \hat\psi(\boldsymbol{x},t)
	=-i\hbar \textcolor{black}{c}\bm{\alpha} \cdot \boldsymbol{\nabla} \hat\psi(\boldsymbol{x},t)
	+\left(mc^2+\hbar^{\frac{1}{2}}c^{\frac{3}{2}}g\phi(\boldsymbol{x},t)\right)\beta \hat\psi(\boldsymbol{x},t) \; , 
\end{equation}
where 
$$    
\alpha_i
=\left(                 
\begin{array}{cc}   
	0 & \sigma_i \\  
	\sigma_i & 0 \\  
\end{array}
\right)                 
\quad,\quad
\beta
=\left(                 
\begin{array}{cc}   
	I & 0\\  
	0 & -I \\  
\end{array}
\right) \; .
$$

The variables are still operators in \eqref{eomope}, whereas in classical theory, all physical quantities are represented by c-numbers, not operators. In the full quantum theory, the state space encompasses the entire Hilbert space, whereas in classical theory, there is no particle creation or annihilation, and only one-particle states evolve. Therefore, it is necessary to reduce the full Hilbert space to a subspace consisting solely of one-particle states, where the operators in the equation of motion can be represented by c-numbers.
However, if the external field $\phi(\boldsymbol{x},t)$ in \eqref{eomope} changes too rapidly with time, there is a finite probability for the one-particle state to transition into a multi-particle state. Hence, it is also important to ensure that $\phi(\boldsymbol{x},t)$ changes relatively slowly with time, so that the probability of the one-particle state evolving into a multi-particle state can be neglected

The wave function for one-particle state $\left|\psi,t\right>=e^{-i\hat H_2 t}\left|\psi\right>$ is defined as \cite{Weinberg2005}
\begin{equation}\label{wave}
	\psi(\boldsymbol{x},t)=\left<\Omega\right|\hat\psi(\boldsymbol{x},0)\left|\psi,t\right>
	=\left<\Omega\right|\hat{\psi}(\boldsymbol{x},0)e^{-i\hat H_2 t}\left|\psi\right>
	=\left<\Omega\right|\hat{\psi}(\boldsymbol{x},t)\left|\psi\right> \; .
\end{equation}

In traditional quantum mechanics, the state $\left|x\right>$ describes a particle located at position $x$, and the wave function of a quantum state $\left|\psi,t\right>$ is the inner product of the state $\left|x\right>$ and $\left|\psi,t\right>$, i.e., $\left<x|\psi,t\right>$.
In quantum field theory, the quantum state $\hat{\psi}^\dagger(\boldsymbol{x},0)\left|\Omega\right>$ can be regarded as a one-particle state located at position $x$ (while $\hat{\psi}(\boldsymbol{x},0)\left|\Omega\right>$ represents the antiparticle state), and \eqref{wave} can be seen as the inner product between $\hat{\psi}^\dagger(\boldsymbol{x},0)\left|\Omega\right>$ and the quantum state $\left|\psi,t\right>=e^{-i\hat H_2 t}\left|\psi\right>$.
Therefore, the definition \eqref{wave} of the wave function is consistent with the physical interpretation of the wave function in traditional quantum mechanics. From the equation of motion \eqref{eomope}, we can obtain a wave function version of the Dirac equation with an external field $\phi(\boldsymbol{x},t)$:
\begin{equation}	\label{dirac1}
	i\hbar\frac{\partial}{\partial t} \psi(\boldsymbol{x},t)
	=c\bm{\alpha} \cdot \hat{\boldsymbol{p}} \psi(\boldsymbol{x},t)
	+\left(mc^2+\hbar^{\frac{1}{2}}c^{\frac{3}{2}}g\phi(\boldsymbol{x},t)\right)\beta \psi(\boldsymbol{x},t)  \; ,
\end{equation}
where $\hat{\boldsymbol{p}}=-i\hbar \boldsymbol{\nabla}$ represents the momentum operator in traditional quantum mechanics. Using \eqref{dirac1} we can express the time derivative of the average position of the wave function as
\begin{equation}	
	\begin{split}
		&\frac{d}{d t}
		\int d^3x\; \psi^\dagger(\boldsymbol{x},t)\boldsymbol{x}\psi(\boldsymbol{x},t)\\
		&=	\int d^3x\; \psi^\dagger(\boldsymbol{x},t)\boldsymbol{x}\frac{\partial}{\partial t}\psi(\boldsymbol{x},t)
		+
		\int d^3x\; \left[\frac{\partial}{\partial t}\psi^\dagger(\boldsymbol{x},t)\right]\boldsymbol{x}\psi(\boldsymbol{x},t)\\
		&=
		\int d^3x\;  \psi^\dagger(\boldsymbol{x},t)
		c\bm{\alpha} \psi(\boldsymbol{x},t)  \; .
	\end{split}
\end{equation}
The above equation indicates that $c\bm{\alpha}$ represents the velocity operator.

From \eqref{dirac1}, we can also express the time derivative of the average momentum of the wave function as
\begin{equation}\label{dpdt}	
	\begin{split}
		&\frac{d}{d t}
		\int d^3x \;\psi^\dagger(\boldsymbol{x},t)\hat{\boldsymbol{p}}\psi(\boldsymbol{x},t)\\
		&=	\int d^3x\; \psi^\dagger(\boldsymbol{x},t)\hat{\boldsymbol{p}}\frac{\partial}{\partial t}\psi(\boldsymbol{x},t)
		+
		\int d^3x\; \left[\frac{\partial}{\partial t}\psi^\dagger(\boldsymbol{x},t)\right]\hat{\boldsymbol{p}}\psi(\boldsymbol{x},t)\\
		&=\frac{1}{i\hbar}
		\left[\hbar^{\frac{1}{2}}c^{\frac{3}{2}}g
		\int d^3x\; \psi^\dagger(\boldsymbol{x},t)\hat{\boldsymbol{p}}
		\phi(\boldsymbol{x},t)\beta \psi(\boldsymbol{x},t)
		-
		\hbar^{\frac{1}{2}}c^{\frac{3}{2}}g
		\int d^3x\; \phi(\boldsymbol{x},t) \psi^\dagger(\boldsymbol{x},t)\beta
		\hat{\boldsymbol{p}}\psi(\boldsymbol{x},t)\right]\\
		&=- \hbar^{\frac{1}{2}}c^{\frac{3}{2}}g
		\int d^3x\; \psi^\dagger(\boldsymbol{x},t)\boldsymbol{\nabla}
		\phi(\boldsymbol{x},t)\beta \psi(\boldsymbol{x},t)  \; .
	\end{split}
\end{equation}
Notably, the point particle in classical theory corresponds to a wave packet in quantum theory, and the extension of the Gaussian wave packet approaches zero in both coordinate space and momentum space in the limit $\hbar\to 0$ \cite{Cohn1972a,Kazandjian2006,Kazandjian2007,Klein2012}.
Therefore, we require the wave function $\psi(\boldsymbol{x},t)$ to be a wave packet. Suppose the peak of $\psi(\boldsymbol{x},t)$ is located at $\boldsymbol{r}_1(t)$; then, in the limit $\hbar \to 0$, Eq. \eqref{dpdt} can be further expressed as follows:
\begin{equation}\label{2f}	
	\begin{split}
		\frac{d}{d t}
		\int d^3x \;\psi^\dagger(\boldsymbol{x},t)\hat{\boldsymbol{p}}\psi(\boldsymbol{x},t)
		&= -\boldsymbol{\nabla}\phi(\boldsymbol{r}_1,t) \hbar^{\frac{1}{2}}c^{\frac{3}{2}}g \int d^3x\; \psi^\dagger(\boldsymbol{x},t)\beta\psi(\boldsymbol{x},t)
		\; .
	\end{split}
\end{equation}

Regarding the meaning of the average value of $\beta$, we can draw upon the expression for the energy of a point particle in free field theory. In quantum mechanics, the Hamiltonian for a free particle is given by $\hat H_0=c\bm{\alpha} \cdot \hat{\boldsymbol{p}}_0+m_1c^2\beta$, then we have
\begin{equation}\label{beta}
	\begin{split}
		&\int d^3x\; \psi^\dagger(\boldsymbol{x},t)\beta \psi(\boldsymbol{x},t)\\
		&= \frac{1}{m_1c^2}\int d^3x\; \psi^\dagger(\boldsymbol{x},t)\hat H_0 \psi(\boldsymbol{x},t)
		-\frac{1}{m_1c^2}\int d^3x\; \psi^\dagger(\boldsymbol{x},t)c\bm{\alpha} \cdot \hat{\boldsymbol{p}}_0\psi(\boldsymbol{x},t)
		\; .
	\end{split}
\end{equation}

In classical theory, the energy and momentum of a particle with velocity $\dot{\boldsymbol{r}}$ are given by $E=\frac{m_1 c^2}{\sqrt{1-\dot{\boldsymbol{r}}^2/c^2}}$ and $\boldsymbol{p}_0=\frac{m_1 \dot{\boldsymbol{r}}}{\sqrt{1-\dot{\boldsymbol{r}}^2/c^2}}$ respectively. Although $\psi(\boldsymbol{x},t)$ is not the wave function of a free particle, the wave packet $\psi(\boldsymbol{x},t)$ is spatially localized around $\boldsymbol{r}_1(t)$, then $\int d^3x\; \psi^\dagger(\boldsymbol{x},t)\hat H_0 \psi(\boldsymbol{x},t)$ in \eqref{beta} can be interpreted as the instantaneous energy of a point particle without potential energy, i.e. $\int d^3x\; \psi^\dagger(\boldsymbol{x},t)\hat H_0 \psi(\boldsymbol{x},t)=\frac{m_1 c^2}{\sqrt{1-\dot{\boldsymbol{r}}_1^2/c^2}}$. Furthermore, since $c\bm{\alpha}$ represents the velocity operator of the particle, based on the properties of the wave packet, it can be inferred that $\int d^3x\; \psi^\dagger(\boldsymbol{x},t)c\bm{\alpha} \cdot \hat{\boldsymbol{p}}_0\psi(\boldsymbol{x},t)=\dot{\boldsymbol{r}}_1\cdot\boldsymbol{p}_0=\frac{m_1 \dot{\boldsymbol{r}}_1^2}{\sqrt{1-\dot{\boldsymbol{r}}_1^2/c^2}}$. By substituting these results into \eqref{beta}, we obtain
\begin{equation}\label{1beta}	
	\begin{split}
		\int d^3x\; \psi^\dagger(\boldsymbol{x},t)\beta \psi(\boldsymbol{x},t)
		= \sqrt{1-\dot{\boldsymbol{r}}_1^2/c^2}
		\; .
	\end{split}
\end{equation}
Denoting the momentum  average of the wave packet as $\boldsymbol{p}_1=\int d^3x \psi^\dagger(\boldsymbol{x},t)\hat{\boldsymbol{p}}\psi(\boldsymbol{x},t)$ and using \eqref{1beta}, we can rewrite  \eqref{2f} as
\begin{equation}\label{3f}	
	\begin{split}
		\frac{d \boldsymbol{p}_1}{dt}
		= -\boldsymbol{\nabla}\phi(\boldsymbol{r}_1,t) \hbar^{\frac{1}{2}}c^{\frac{3}{2}}g \sqrt{1-\dot{\boldsymbol{r}}_1^2/c^2}   
		\; .
	\end{split}
\end{equation}

For a Lagrangian $L$, the Euler-Lagrange equations are given by $\frac{d p^i}{dt}=\frac{d}{dt}\left(\frac{\partial L}{\partial \dot{r}^i}\right)=\frac{\partial L}{\partial r^i}$. Therefore, we can deduce the Lagrangian corresponding to the equations of motion \eqref{3f} as follows
\begin{equation}\label{4f}	
	\begin{split}
		L_2(\boldsymbol{r}_1,\dot{\boldsymbol{r}_1},t)=-\hbar^{\frac{1}{2}}c^{\frac{3}{2}}g\phi(\boldsymbol{r}_1,t)\sqrt{1-\dot{\boldsymbol{r}_1}^2/c^2}+f(\dot{\boldsymbol{r}}_1,t)
		\; .
	\end{split}
\end{equation}

From Equation \eqref{eomope}, it is evident that when the external field $\phi(\boldsymbol{r}_1,t)=0$, the system returns to the scenario of a completely free field. Thus, when $\phi(\boldsymbol{r}_1,t)=0$, \eqref{4f} yields the Lagrangian for a free particle, given by $L(\boldsymbol{r}_1,\dot{\boldsymbol{r}_1},t)=f(\dot{\boldsymbol{r}}_1,t)=-m_1 c^2\sqrt{1-\dot{\boldsymbol{r}_1}^2/c^2}$. Therefore, \eqref{4f} can be written as
\begin{equation}\label{5f}	
	\begin{split}
		L_2(\boldsymbol{r}_1,\dot{\boldsymbol{r}_1},t)=-\left(m_1 c^2+\hbar^{\frac{1}{2}}c^{\frac{3}{2}}g\phi(\boldsymbol{r}_1,t)\right)\sqrt{1-\dot{\boldsymbol{r}_1}^2/c^2}
		\; .
	\end{split}
\end{equation}
This provides the classical correspondence of the quantum field theory described by the Hamiltonian $\hat{H}_2$ in \eqref{spllag}. 
The classical theory described by the Lagrangian \eqref{5f} explains the motion of a point particle in an external field $\phi(\boldsymbol{x},t)$,
and its Euler-Lagrange equation plays a role in the classical Yukawa theory analogous to the Lorentz force in electrodynamics.

Summarizing the results of previous sections, the Lagrangians \eqref{L1} and \eqref{5f} together constitute the relativistic classical Yukawa theory.

\section{Relativistic corrections to the Yukawa potential}
\label{sec:Yukawa}
In this section, we will derive the relativistic corrections to the Yukawa potential.
We first review the relativistic corrections to the Coulomb potential \cite{Darwin1920}. In electrodynamics, the classical Lagrangian for a charged particle in an external electromagnetic field can be written as
\begin{equation}\label{Leq1}
	L_{\text{ED}}=-m_1\textcolor{black}{c^2}\sqrt{1-\frac{\dot{\boldsymbol{r}}_1^2}{\textcolor{black}{c^2}}}-e_1 A^0(\boldsymbol{r}_1,t) +\frac{e_1}{\textcolor{black}{c}}\dot{\boldsymbol{r}}_1\cdot\boldsymbol{A}(\boldsymbol{r}_1,t)
	\; ,
\end{equation}
where $m_1$ is the particle mass and $\boldsymbol{r}_1$ is the position vector of the charged particle. The subscript ED represents electrodynamics. Another Lagrangian that can give Maxwell's equations is
\begin{equation}
	\begin{split}
		\mathcal{L}_{\text{ED}}=-\frac{1}{4}F^{\mu\nu}F_{\mu\nu}-e_2j^{\mu}A_{\mu}
		\; .
	\end{split}
\end{equation}

Now we consider another particle with a trajectory $\boldsymbol{r}_2(t_2)$, The Lienard-Wiechert potential of the particle can be derived from Lagrangian $\mathcal{L}_{ED}$ as
\begin{equation}
	\begin{split}\label{A}
		&{\color{black}A^0} =\frac{e_2}{|\boldsymbol{r}_2(t^*)-\boldsymbol{r}_1|+\dot{\boldsymbol{r}}_2(t^*)\cdot\left(\boldsymbol{r}_2(t^*)-\boldsymbol{r}_1\right)/\textcolor{black}{c}} \; ,\\
		&\boldsymbol{A}=\frac{e_2}{c}\frac{\dot{\boldsymbol{r}}_2(t^*)}{|\boldsymbol{r}_2(t^*)-\boldsymbol{r}_1|+\dot{\boldsymbol{r}}_2(t^*)\cdot\left(\boldsymbol{r}_2(t^*)-\boldsymbol{r}_1\right)/\textcolor{black}{c}}
		\; ,
	\end{split}	
\end{equation}
where $t_2^*$ satisfies $t_2^*=t-\left|\boldsymbol{r}_1-\boldsymbol{r}_2(t_2^*)\right|/c$ and $e_2$ is the charge of the particle. In order to obtain the effective potential of the two charged particles, we need to expree \eqref{A} in terms of $\boldsymbol{r}_2(t)$ instead of $\boldsymbol{r}_2(t^*)$. The difference between $t$ and $t_2^*$ can be written as
\begin{equation}
	\label{tau}
	t-t_2^*=\frac{r}{c}-\frac{\dot{\boldsymbol{r}}_2\cdot\left(\boldsymbol{r}_2-\boldsymbol{r}_1\right)}{\textcolor{black}{c^2}}
	+\frac{r}{2\textcolor{black}{c^3}}\times\left[\dot{\boldsymbol{r}}_2^2+\ddot{\boldsymbol{r}}_2\cdot\left(\boldsymbol{r}_2-\boldsymbol{r}_1\right)+\frac{\left(\dot{\boldsymbol{r}}_2\cdot\left(\boldsymbol{r}_2-\boldsymbol{r}_1\right)\right)^2}{r^2}\right]
	\; ,
\end{equation}
where  $\boldsymbol{r}_2=\boldsymbol{r}_2(t)$, $\dot{\boldsymbol{r}}_2=\dot{\boldsymbol{r}}_2(t)$, $\ddot{\boldsymbol{r}}_2=\ddot{\boldsymbol{r}}_2(t)$, and $r=|\boldsymbol{r}_2-\boldsymbol{r}_1|$. All the variables mentioned above depend only on the time $t$. 
By combining \eqref{tau}, \eqref{A}and\eqref{Leq1}, C.Darwin derived the Lagrangian for two charged particles \cite{Darwin1920} 
\begin{equation}\label{qedcor}
	\begin{split}
		L_t=-m_1\textcolor{black}{c^2}\sqrt{1-\frac{\dot{\boldsymbol{r}}_1^2}{\textcolor{black}{c^2}}}&-m_2\textcolor{black}{c^2}\sqrt{1-\frac{\dot{\boldsymbol{r}}_2^2}{\textcolor{black}{c^2}}}-\frac{e_1e_2}{r}\\
		&+\frac{e_1e_2}{2\textcolor{black}{c^2}}\left[\frac{\dot{\boldsymbol{r}}_1\cdot\dot{\boldsymbol{r}}_2}{r}+\frac{\dot{\boldsymbol{r}}_1\cdot\left(\boldsymbol{r}_2-\boldsymbol{r}_1\right)\dot{\boldsymbol{r}}_2\cdot\left(\boldsymbol{r}_2-\boldsymbol{r}_1\right)}{r^3}\right]
		\; ,
	\end{split}	
\end{equation}
where the second line represents the relativistic corrections to the Coulomb potential.

Now that we have the Lagrangians for classical Yukawa theory, namely \eqref{L1} and \eqref{5f}, we can proceed to derive the relativistic corrections to the Yukawa potential using a similar procedure as in electrodynamics. By considering Lagrangians \eqref{L1} and \eqref{5f}, we obtain the actions for the entire classical Yukawa theory
\begin{equation}\label{S}
	\begin{split}
		&S_1=\int d^4x\left(\frac{1}{2}\partial_{\mu}\phi\partial^{\mu}\phi-\frac{1}{2}\left(\frac{m_0c^2}{\hbar}\right)^2\phi^2-\hbar^{\frac{1}{2}}c^{\frac{3}{2}}gj\phi\right)\\
		&S_2=-\int dt\left(m_1c^2+\hbar^{\frac{1}{2}}c^{\frac{3}{2}}g\phi\right)\sqrt{1-\dot{\boldsymbol{r}}^2/c^2}
		\; .
	\end{split}
\end{equation} 
Here, $S_1$ describes the time evolution of the classical field $\phi$ in the presence of an external source $j$, while $S_2$ describes the motion of a particle influenced by an external field $\phi$.
By comparing the forms of these two actions, we can know that the external source is $j=\sqrt{1-\dot{\boldsymbol{r}}_2^2/c^2}\delta(\boldsymbol{r}-\boldsymbol{r}(t))$ for a Yukawa-type charged particle with the trajectory $\boldsymbol{r}(t)$. In order to study the interaction between two Yukawa-type charged particles, we derive the equations of motion from the actions \eqref{S}:
\begin{equation}\label{E2}
	-\ddot{\phi}(\boldsymbol{r}_1,t_1)+c^2\nabla_1^2\phi(\boldsymbol{r}_1,t_1)-\left(\frac{m_0c^2}{\hbar}\right)^2\phi(\boldsymbol{r}_1,t_1)=\hbar^{\frac{1}{2}}c^{\frac{3}{2}}g_2\sqrt{1-\dot{\boldsymbol{r}}_2^2(t_1)/c^2}\delta(\boldsymbol{r}_1-\boldsymbol{r}_2(t_1))
\end{equation}
\begin{equation}\label{E1}
	\frac{d}{dt}\left[\left(m_1c^2+\hbar^{\frac{1}{2}}c^{\frac{3}{2}}g\phi(\boldsymbol{r}_1,t_1)\right)\frac{\dot{\boldsymbol{r}}_1/c^2}{\sqrt{1-\dot{\boldsymbol{r}}_1^2/c^2}}\right]=-\hbar^{\frac{1}{2}}c^{\frac{3}{2}}g_1\sqrt{1-\dot{\boldsymbol{r}}_1^2/c^2}\nabla_1\phi(\boldsymbol{r}_1,t_1)
\end{equation}
where  $\boldsymbol{r}_2(t)$ represents the trajectory of the source particle, $\boldsymbol{r}_1$ is the position vector of the affected particle, and $g_2$ and $g_1$ represent the Yukawa-type charges of these two particles, respectively. The equation of motion for the bosonic field \eqref{E2} is evidently a wave equation with a source term $g_2\sqrt{1-\dot{\boldsymbol{r}}_2^2(t_1)/c^2}\delta(\boldsymbol{r}_1-\boldsymbol{r}_2(t_1))$. For such a wave equation, it is convenient to express the solution in terms of a Green function
\begin{equation}\label{phigreen}
	\phi(\boldsymbol{r}_1,t_1)=\int dt_2 \; \hbar^{\frac{1}{2}}c^{\frac{3}{2}}g_2\sqrt{1-\dot{\boldsymbol{r}}^2_2(t_2)/c^2}G\left(\boldsymbol{r}_1,\boldsymbol{r}_2(t_2);t_1,t_2\right)
	\; ,
\end{equation}
where $G\left(\boldsymbol{r}_1,\boldsymbol{r}_2;t_1,t_2\right)$ is the Green function satisfying the Green equation
\begin{equation}
	\begin{split}
		-\ddot{G}\left(\boldsymbol{r}_1,\boldsymbol{r}_2;t_1,t_2\right)&+c^2\nabla_1^2G\left(\boldsymbol{r}_1,\boldsymbol{r}_2;t_1,t_2\right)-\left(\frac{m_0c^2}{\hbar}\right)^2G\left(\boldsymbol{r}_1,\boldsymbol{r}_2;t_1,t_2\right)=\delta\left(\boldsymbol{r}_1-\boldsymbol{r}_2\right)\delta\left(t_1-t_2\right)
		\; .
	\end{split}	
\end{equation}

Obviously, the desired Green function is of the causal retarded type. It can be expressed as
\begin{equation}
	\begin{split}\label{Green}
		G\left(\boldsymbol{r}_1,\boldsymbol{r}_2;t_1,t_2\right)
		&=
		\theta(t_1-t_2)
		\theta\left(\left(t_2-t_1\right)^2-\left(\boldsymbol{r}_1-\boldsymbol{r}_2\right)^2/c^2\right)\\
		&\times
		\frac{m_0}{4\pi\hbar c \sqrt{\left(t_2-t_1\right)^2-\left(\boldsymbol{r}_1-\boldsymbol{r}_2\right)^2/c^2}}J_1\left(\frac{m_0c^2\sqrt{\left(t_2-t_1\right)^2-\left(\boldsymbol{r}_1-\boldsymbol{r}_2\right)^2/c^2}}{\hbar}\right)\\
		&-\theta(t_1-t_2)\frac{1}{2\pi c^3}\delta\left(\left(t_2-t_1\right)^2-\left(\boldsymbol{r}_1-\boldsymbol{r}_2\right)^2/c^2\right)
		\; ,
	\end{split}	
\end{equation}	
where $\theta(t)$ is the Heaviside step function. It is important to note that  $\left(t_2-t_1\right)^2-\left(\boldsymbol{r}_1-\boldsymbol{r}_2\right)^2/c^2=0$ is a critical point. And we introduce $t_2^*$ defined as $t_2^*=t_1-\left|\boldsymbol{r}_1-\boldsymbol{r}_2(t_2^*)\right|/c$, which satisfies $\left(t^*_2-t_1\right)^2-\left(\boldsymbol{r}_1-\boldsymbol{r}_2(t^*_2)\right)^2/c^2=0$  and $t^*_2<t_1$ simultaneously.
For convenience, we also introduce the variables  $\kappa=1/c$, $\tau=c(t_2-t_1)$ and $\tau^*=c(t_2^*-t_1)$. In order to solve the integral in \eqref{phigreen}, we define a new integral variable
\begin{equation}	
	y=\left\{\begin{array}{ll}		
		\sqrt{\tau^2-\left[\boldsymbol{r}_1-\boldsymbol{r}_2(t_1+\kappa\tau)\right]^2}&,\quad\tau\leqslant\tau^*\\
		-\sqrt{\tau^2-\left[\boldsymbol{r}_1-\boldsymbol{r}_2(t_1+\kappa\tau)\right]^2}&,\quad\tau\geqslant\tau^*
	\end{array}\right.
	\; .
\end{equation}
Substituting the Green function \eqref{Green} into \eqref{phigreen} and using the new variable $y$, we can rewrite \eqref{phigreen} as
\begin{equation}
	\begin{split}
		\label{phi1}
		\phi\left(\boldsymbol{r}_1,t_1\right)
		=&-\frac{m_0\hbar^{\frac{1}{2}}c^{\frac{3}{2}}g_2}{4\pi\hbar c}\int_{0}^{\infty} dy \frac{d\tau}{dy}\sqrt{1-\kappa^2\dot{\boldsymbol{r}}_2^2(t_1+\kappa\tau)}\frac{J_1\left(\frac{m_0cy}{\hbar}\right)}{y}\\
		&\;\;\;\;\;\;+\frac{\hbar^{\frac{1}{2}}c^{\frac{3}{2}}g_2}{4\pi c^2}
		\int_{-\infty}^{\infty}  dy 
		\frac{d\tau}{dy} \sqrt{1-\kappa^2\dot{\boldsymbol{r}}_2^2(t_1+\kappa\tau)}
		\frac{\delta\left(y\right)}{|y|}
		\; .
	\end{split}
\end{equation}
If we consider the relativistic corrections only up to $\mathcal{O}(1/c^2)$, then $\tau$ can be approximated as
\begin{equation}
	\begin{split}
		\tau&=\tau\bigg|_{\kappa=0}+\kappa\frac{\partial\tau}{\partial\kappa}\bigg|_{\kappa=0}
		+\frac{1}{2}\kappa^2\frac{\partial^2\tau}{\partial\kappa^2}\bigg|_{\kappa=0}
		\; .
	\end{split}	
\end{equation}
Denoting $\boldsymbol{r}_2=\boldsymbol{r}_2(t_1)$ and $\boldsymbol{r}=\boldsymbol{r}_1-\boldsymbol{r}_2$,  we can express the partial derivative of $\tau$ with respect to $\kappa$ at $\kappa=0$ as
\begin{equation}
	\begin{split}
		\frac{\partial\tau}{\partial\kappa}\bigg|_{\kappa=0}
		=-\frac{\;\;\frac{\partial y}{\partial\kappa}\;\;}{\;\;\frac{\partial y}{\partial\tau}\;\;}\bigg|_{\kappa=0}
		=\dot{\boldsymbol{r}}_2\cdot\boldsymbol{r}
		\; ,
	\end{split}	
\end{equation}
and the second derivative of $\tau$ with respect to $\kappa$ at $\kappa=0$ as
\begin{equation}
	\begin{split}
		\frac{\partial^2\tau}{\partial\kappa^2}\bigg|_{\kappa=0}
		&	=\frac{-\frac{\partial^2 y}{\partial\kappa^2}\frac{\partial y}{\partial\tau}
			+2\frac{\partial^2 y}{\partial\kappa\partial\tau}\frac{\partial y}{\partial\kappa}
			+\frac{\partial^2 y}{\partial\tau^2}\frac{\partial y}{\partial\kappa}
			\frac{\partial \tau}{\partial\kappa}}
		{\left(\frac{\partial y}{\partial\tau}\right)^2}\bigg|_{\kappa=0}\\
		&=
		-\frac{1}{\sqrt{y^2+\boldsymbol{r}^2}}\left[\dot{\boldsymbol{r}}^2_2\left(y^2+\boldsymbol{r}^2\right)+\left(\dot{\boldsymbol{r}}_2\cdot\boldsymbol{r}\right)^2-\ddot{\boldsymbol{r}}_2\cdot\boldsymbol{r}\left(y^2+\boldsymbol{r}^2\right)\right]
		\; .
	\end{split}	
\end{equation}
Thus, the term $\frac{d\tau}{dy}$ in the integral \eqref{phi1} can be written as
\begin{equation}	
	\label{ex1}
	\frac{d\tau}{dy}=
	-\left(1+\frac{\dot{\boldsymbol{r}}^2_2-\ddot{\boldsymbol{r}}_2\cdot\boldsymbol{r}}{2c^2}\right)
	\frac{|y|}{\sqrt{y^2+\boldsymbol{r}^2}}			
	+\frac{\left(\dot{\boldsymbol{r}}_2\cdot\boldsymbol{r}\right)^2
	}{2c^2}\frac{|y|}{\left(y^2+\boldsymbol{r}^2\right)^\frac{3}{2}}
	\; .
\end{equation}

On the other hand, we can expand another term in the integral\eqref{phi1} as
\begin{equation}\label{ex2}
	\sqrt{1-\kappa^2\dot{\boldsymbol{r}}_2^2(t_1+\kappa\tau)}=1-\frac{1}{2}\dot{\boldsymbol{r}}_2^2\kappa^2
	\; .
\end{equation}
By substituting the expansion \eqref{ex1} and \eqref{ex2} into \eqref{phi1}, we can now perform the integral
\begin{equation}\label{soleomcor}
	\begin{split}
		\phi\left(\boldsymbol{r}_1,t_1\right)
		&=\frac{m_0\hbar^{\frac{1}{2}}c^{\frac{3}{2}}g_2}{4\pi\hbar c}\left[ \left(1-\frac{\ddot{\boldsymbol{r}}_2\cdot\boldsymbol{r}}{2c^2}\right)
		\int_{0}^{\infty} dy\frac{J_1\left(\frac{m_0cy}{\hbar}\right)}
		{\sqrt{y^2+\boldsymbol{r}^2}}			
		-\frac{\left(\dot{\boldsymbol{r}}_2\cdot\boldsymbol{r}\right)^2
		}{2c^2}
		\int_{0}^{\infty} dy\frac{J_1\left(\frac{m_0cy}{\hbar}\right)}
		{\left(y^2+\boldsymbol{r}^2\right)^\frac{3}{2}}\right]\\
		&\quad-\frac{\hbar^{\frac{1}{2}}c^{\frac{3}{2}}g_2}{4\pi c^2}
		\int_{-\infty}^{\infty}  dy 
		\left[\left(1-\frac{\ddot{\boldsymbol{r}}_2\cdot\boldsymbol{r}}{2c^2}\right)
		\frac{1}{\sqrt{y^2+\boldsymbol{r}^2}}			
		-\frac{\left(\dot{\boldsymbol{r}}_2\cdot\boldsymbol{r}\right)^2
		}{2c^2}\frac{1}{\left(y^2+\boldsymbol{r}^2\right)^\frac{3}{2}}\right]
		\delta\left(y\right)\\
		&
		=\frac{\hbar^{\frac{1}{2}}c^{\frac{3}{2}}g_2}{4\pi c^2}\left[-\frac{e^{-\frac{m_0c^2}{\hbar c}r}}{r}
		+\frac{\ddot{\boldsymbol{r}}_2\cdot\boldsymbol{r}}{2c^2}\frac{e^{-\frac{m_0c^2}{\hbar c}r}}{r}
		+\frac{\left(\dot{\boldsymbol{r}}_2\cdot\boldsymbol{r}\right)^2}{2c^2}\left(\frac{1}{r}+\frac{m_0c^2}{\hbar c}\right)\frac{e^{-\frac{m_0c^2}{\hbar c}r}}{r^2}\right]\\
		&\quad+\frac{m_0\hbar^{\frac{1}{2}}c^{\frac{3}{2}}g_2}{4\pi\hbar c}\left[ \left(1-\frac{\ddot{\boldsymbol{r}}_2\cdot\boldsymbol{r}}{2c^2}\right) 	 
		\frac{1}{\frac{m_0c}{\hbar}r}			
		-\frac{\left(\dot{\boldsymbol{r}}_2\cdot\boldsymbol{r}\right)^2
		}{2c^2}
		\frac{1}{\frac{m_0c}{\hbar}r^3}\right]\\
		&\qquad-\frac{\hbar^{\frac{1}{2}}c^{\frac{3}{2}}g_2}{4\pi c^2}
		\left[\left(1-\frac{\ddot{\boldsymbol{r}}_2\cdot\boldsymbol{r}}{2c^2}\right)
		\frac{1}{r}			
		-\frac{\left(\dot{\boldsymbol{r}}_2\cdot\boldsymbol{r}\right)^2
		}{2c^2}\frac{1}{r^3}\right]
		\\
		&=\frac{\hbar^{\frac{1}{2}}c^{\frac{3}{2}}g_2}{4\pi c^2}\left[-\frac{e^{-\frac{m_0c^2}{\hbar c}r}}{r}
		+\frac{\ddot{\boldsymbol{r}}_2\cdot\boldsymbol{r}}{2c^2}\frac{e^{-\frac{m_0c^2}{\hbar c}r}}{r}
		+\frac{\left(\dot{\boldsymbol{r}}_2\cdot\boldsymbol{r}\right)^2}{2c^2}\left(\frac{1}{r}+\frac{m_0c^2}{\hbar c}\right)\frac{e^{-\frac{m_0c^2}{\hbar c}r}}{r^2}\right]
		\; .
	\end{split}
\end{equation}
It should be noted that the classical field $\phi(\boldsymbol{r}_1)$ created by a particle with the trajectory $\boldsymbol{r}_2(t_2)$ depends on $\boldsymbol{r}_1$ only through the relative coordinate $\boldsymbol{r}_1-\boldsymbol{r}_2$, which is a manifestation of translation symmetry. In other words, the classical field $\phi(\boldsymbol{r}_1)$ created by a particle with the trajectory $\boldsymbol{r}_2(t_2)$ remains invariant if both the source particle at $\boldsymbol{r}_2(t_2)$ and the field point $\boldsymbol{r}_1$ are shifted by the same distance $\boldsymbol{r}_0$.

Guided by the C.Darwin's method in \cite{Darwin1920}, we can derive the Lagrangian for two particles from  \eqref{soleomcor} and \eqref{S} as follows
\begin{equation}
	\begin{split}\label{Lyukawa}
		L=&-m_1\sqrt{1-\dot{\boldsymbol{r}}_1^2/\textcolor{black}{c^2}}-m_2\sqrt{1-\dot{\boldsymbol{r}}_2^2/\textcolor{black}{c^2}}\\
		&+\frac{\textcolor{black}{\hbar c}g_1 g_2}{4\pi}
		\frac{e^{-\frac{m_0\textcolor{black}{c^2}}{\textcolor{black}{\hbar c}}r}}{r}
		+\frac{1}{\textcolor{black}{c^2}}\frac{\textcolor{black}{\hbar c}g_1g_2}{8\pi}\left[\frac{\dot{\boldsymbol{r}}_1\cdot\dot{\boldsymbol{r}}_2}{r}
		-\frac{\dot{\boldsymbol{r}}_1^2+\dot{\boldsymbol{r}}_2^2}{r}
		-\left(\frac{1}{r}+\frac{m_0\textcolor{black}{c^2}}{\textcolor{black}{\hbar c}}\right)\frac{\left(\dot{\boldsymbol{r}}_1\cdot\boldsymbol{r}\right)\left(\dot{\boldsymbol{r}}_2\cdot\boldsymbol{r}\right)}{r^2}\right]e^{-\frac{m_0\textcolor{black}{c^2}}{\textcolor{black}{\hbar c}}r}
		\; ,
	\end{split}	
\end{equation}
where $m_2$ is the mass of the particle with the trajectory $\boldsymbol{r}_2(t)$.	
From the Lagrangian \eqref{Lyukawa}, we can derive the Hamiltonian of the system
\begin{equation}\label{Hyukawa}
	\begin{split}
		H&=\boldsymbol{p}_1\cdot\dot{\boldsymbol{r}}_1+\boldsymbol{p}_2\cdot\dot{\boldsymbol{r}}_2
		-L\\
		&=\frac{m_1 c^2}{\sqrt{1-\dot{\boldsymbol{r}}_1^2/\textcolor{black}{c^2}}}
		+\frac{m_2 c^2}{\sqrt{1-\dot{\boldsymbol{r}}_2^2/\textcolor{black}{c^2}}}
		-\frac{\textcolor{black}{\hbar c}g_1g_2}{4\pi}
		\frac{e^{-\frac{m_0\textcolor{black}{c^2}}{\textcolor{black}{\hbar c}}r}}{r}\\
		&+\frac{1}{\textcolor{black}{c^2}}\frac{\textcolor{black}{\hbar c}g_1g_2}{8\pi}\left[\frac{\dot{\boldsymbol{r}}_1\cdot\dot{\boldsymbol{r}}_2}{r}
		-\frac{\dot{\boldsymbol{r}}_1^2+\dot{\boldsymbol{r}}_2^2}{r}
		-\left(\frac{1}{r}+\frac{m_0\textcolor{black}{c^2}}{\textcolor{black}{\hbar c}}\right)\frac{\left(\dot{\boldsymbol{r}}_1\cdot\boldsymbol{r}\right)\left(\dot{\boldsymbol{r}}_2\cdot\boldsymbol{r}\right)}{r^2}\right]e^{-\frac{m_0\textcolor{black}{c^2}}{\textcolor{black}{\hbar c}}r}\\
		&=\sqrt{\boldsymbol{p}_1^2c^2+m_1^2c^4}+\sqrt{\boldsymbol{p}_1^2c^2+m_1^2c^4}
		-\frac{\textcolor{black}{\hbar c}g_1g_2}{4\pi}
		\frac{e^{-\frac{m_0\textcolor{black}{c^2}}{\textcolor{black}{\hbar c}}r}}{r}\\
		&-\frac{1}{\textcolor{black}{c^2}}\frac{\textcolor{black}{\hbar c}g_1g_2}{8\pi}
		\left[\frac{\boldsymbol{p}_1\cdot\boldsymbol{p}_2}{m_1 m_2}
		-\frac{\boldsymbol{p}_1^2}{m_1^2}	-\frac{\boldsymbol{p}_2^2}{m_2^2}
		-\frac{1}{m_1 m_2}\left(1+\frac{m_0\textcolor{black}{c^2}}{\textcolor{black}{\hbar c}}r\right)\frac{\left(\boldsymbol{p}_1\cdot\boldsymbol{r}\right)\left(\boldsymbol{p}_2\cdot\boldsymbol{r}\right)}{r^2}\right]\frac{e^{-\frac{m_0\textcolor{black}{c^2}}{\textcolor{black}{\hbar c}}r}}{r}\\
	\end{split}	
\end{equation}
Hence, we can read off the potential between two moving Yukawa-charged particles
\begin{equation}\label{V}
	\begin{split}
		V=&-\frac{\textcolor{black}{\hbar c}g_1g_2}{4\pi}
		\frac{e^{-\frac{m_0\textcolor{black}{c^2}}{\textcolor{black}{\hbar c}}r}}{r}
		-\frac{1}{\textcolor{black}{c^2}}\frac{\textcolor{black}{\hbar c}g_1g_2}{8\pi}
		\left[
		-\frac{\boldsymbol{p}_1^2}{m_1^2}	-\frac{\boldsymbol{p}_2^2}{m_2^2}
		\right]\frac{e^{-\frac{m_0\textcolor{black}{c^2}}{\textcolor{black}{\hbar c}}r}}{r}\\
		&\qquad-\frac{1}{\textcolor{black}{c^2}}\frac{\textcolor{black}{\hbar c}g_1g_2}{8\pi}
		\left[\frac{\boldsymbol{p}_1\cdot\boldsymbol{p}_2}{m_1 m_2}
		-\frac{1}{m_1 m_2}\left(1+\frac{m_0\textcolor{black}{c^2}}{\textcolor{black}{\hbar c}}r\right)\frac{\left(\boldsymbol{p}_1\cdot\boldsymbol{r}\right)\left(\boldsymbol{p}_2\cdot\boldsymbol{r}\right)}{r^2}\right]\frac{e^{-\frac{m_0\textcolor{black}{c^2}}{\textcolor{black}{\hbar c}}r}}{r}\\
		=&V_0(r)
		-\frac{1}{c^2}\frac{1}{2m_1^2}\boldsymbol{p}_1V_0(r)\cdot\boldsymbol{p}_1	-\frac{1}{c^2}\frac{1}{2m_2^2}\boldsymbol{p}_2V_0(r)\cdot\boldsymbol{p}_2\\
		&\qquad\;+\frac{1}{c^2}\frac{1}{2m_1 m_2}
		\left[\boldsymbol{p}_1V_0(r)\cdot\boldsymbol{p}_2
		+
		\boldsymbol{p}_1\cdot\boldsymbol{r} \frac{V'_0(r)}{r}\boldsymbol{r}\cdot\boldsymbol{p}_2\right]
		\; ,
	\end{split}	
\end{equation}
where $V'_0(r)\equiv\frac{d}{dr}V_0(r)$. 
$V_0(r)=-\frac{\textcolor{black}{\hbar c}g_1g_2}{4\pi}
\frac{e^{-\frac{m_0\textcolor{black}{c^2}}{\textcolor{black}{\hbar c}}r}}{r}$ is the traditional Yukawa potential describing interaction between nucleons\cite{Yukawa:1935xg}, while the other terms that depend on momentum are the relativistic corrections to the Yukawa potential.

The massless limit $m_0\rightarrow0$ reduces the traditional Yukawa potential to the Coulomb potential. However, in the massless limit, equation \eqref{V} does not converge to C. Darwin's result for electrodynamics, as shown in equation \eqref{qedcor}. This indicates that the relativistic corrections to the interaction through a scalar field is different from the relativistic corrections to the interaction through a vector field. It also implies that a simple extrapolation of non-relativistic potentials may not yield the correct relativistic corrections. Our results, given by equations \eqref{Lyukawa}, \eqref{Hyukawa}, and \eqref{V}, are consistent with results obtained using various other methods \cite{Breit1937, Havas1969, Foldy1961, Woodcock1972, Kennedy1972}. 
This demonstrates the reliability of our approach and results in addressing the classical limit of the Yukawa theory.

\section{conclusions and discussions}	\label{sec:dis}

We have derived the quantum states corresponding to classical fields in the representation expanded by the eigenstates of quantum field operators, and investigated the evolution of such states through the Schrödinger equation, which ultimately yields the evolution equations for the classical fields.
Regarding fermions, classical theory does not include fermionic fields; instead, fermions are treated as point particles. This implies that we only need to address the classical limit of one-particle quantum mechanics, which has already been extensively studied in earlier works.
Finally, we obtain a relativistic classical Yukawa theory, from which we calculate the relativistic corrections to the traditional Yukawa potential.

Although the Yukawa potential with relativistic corrections is derived in classical theory, it can be applied in quantum mechanics as the potential of the Hamiltonian \cite{Breit1937,Childers1987}, similar to the application of Darwin's relativistic corrections in quantum mechanics \cite{Breit1929,Saue2011,DeSanctis2009}.
Yukawa theory is widely used to describe the interactions between fermions through a scalar mediator. For example, in dark phenomenology, self-interacting dark matter with a Yukawa potential appears to provide a better explanation for observed galactic structures compared to collisionless models \cite{Loeb2011}.
To better describe the behavior of dark matter, \cite{Biondini:2021ccr} utilized non-relativistic effective field theory (NREFT) to calculate momentum-dependent effective potentials.
Using Gromes bracket notation\cite{Gromes1977}, the quantum version of the effective potential \eqref{V} can be expressed as follows
\begin{equation}\label{V2}
	\begin{split}
		V=&V_0(r)
		-\frac{1}{c^2}\frac{1}{2m_1^2}
		\left\{\left\{\boldsymbol{p}_1V_0(r)\cdot\boldsymbol{p}_1\right\}\right\}	-\frac{1}{c^2}\frac{1}{2m_2^2}
		\left\{\left\{\boldsymbol{p}_2V_0(r)\cdot\boldsymbol{p}_2\right\}\right\}\\
		&\qquad\;+\frac{1}{c^2}\frac{1}{2m_1 m_2}
		\left[\left\{\left\{\boldsymbol{p}_1V_0(r)\cdot\boldsymbol{p}_2\right\}\right\}
		+
		\left\{\left\{\boldsymbol{p}_1\cdot\boldsymbol{r} \frac{V'_0(r)}{r}\boldsymbol{r}\cdot\boldsymbol{p}_2\right\}\right\}\right]
		\; ,
	\end{split}	
\end{equation}
where Gromes bracket notation$\{\{\}\}$ is defined as
$\left\{\left\{p_iF(r)q_j\right\}\right\}
=\frac{1}{4}\Big[p_iF(r)q_j+q_jF(r)p_i+p_iq_jF(r)+F(r)p_iq_j\Big]$ .	  
By referring to the derivation method in reference \cite{Barnes1982}, it can be observed that the first line of equation \eqref{V2} corresponds precisely to the result obtained from non-relativistic effective quantum field theory. In other words, our result extends beyond the outcome derived from NREFT by including the additional term $V_r\equiv\frac{1}{c^2}\frac{1}{2m_1 m_2}
\left[\left\{\left\{\boldsymbol{p}_1V_0(r)\cdot\boldsymbol{p}_2\right\}\right\}
+
\left\{\left\{\boldsymbol{p}_1\cdot\boldsymbol{r} \frac{V'_0(r)}{r}\boldsymbol{r}\cdot\boldsymbol{p}_2\right\}\right\}\right]$. This discrepancy is an interesting issue. For a stable potential, its relativistic corrections to the Hamiltonian can be obtained, known as the generalized Breit-Fermi Hamiltonian. These relativistic corrections can be divided into spin-dependent and spin-independent components \cite{Lucha1991}, where the "spin" refers to the fermion spin. The spin-independent part, particularly when the interaction is described by a scalar field, has been the subject of significant controversy \cite{Gromes1977,Olsson1983,Barnes1982,McClary1983,Gesztesy1984,Lucha1991,Bhatt1991,Olson1992}.
The effective potential \eqref{V2} is consistent with the results of \cite{Olsson1983}, while the results of NREFT \cite{Biondini:2021ccr} align with \cite{Barnes1982}. The term $V_r$, which accounts for the difference, has later been referred to as the retardation correction or retardation term. It plays a significant role when studying bound-state problems \cite{Bhatt1991,Olson1992}.

The key to deriving the classical limit is to find the quantum states corresponding to the classical fields.
We can outline some natural requirements for quantum states that have classical correspondence. First, vacuum states in quantum theory correspond to the classical field configuration $\phi_\text{class}=0,\pi_\text{class}=0$. Second, if a certain quantum state correspond to a classical field configuration $\phi_\text{class}=\phi_0,\pi_\text{class}=\pi_0$ , then the probability of obtaining $\phi_0,\pi_0$ is the highest when measuring $\hat\phi,\hat\pi$ of the quantum state (in fact, as shown in \ref{subsec:sys}, $\phi_0$ and $\pi_0$ also represent the center of the wave packet), while the probability of obtaining the fields far from $\phi_0,\pi_0$ is much smaller, similar to the behavior of traditional quantum mechanics wave packets that describe classical point particles with definite position and momentum. Finally, a quantum state that exhibits classical field correspondence will continue to maintain this correspondence during its time evolution; it will not evolve into a quantum state that lacks classical correspondence. By analyzing the structure of the vacuum state in the $\phi$/$\pi$-representation and combining the aforementioned constraints, we can find the quantum states that correspond to the classical fields. 
In fact, these quantum states are precisely coherent states. Once we establish the correspondence between quantum states and classical fields, we can determine how the classical fields evolve over time by studying the evolution of the quantum states according to the Schrödinger equation, thereby obtaining the corresponding classical theory.

	In real spacetime representation, the quantum states corresponding to classical fields not only allow us to derive the relativistic classical Yukawa theory but also provide a concrete illustration of the fundamental difference between classical and quantum superposition, even if the double-slit interference patterns produced by these two types of superposition are identical.	
	Furthermore, in addition to the previously mentioned properties, we have found another very important property of the quantum states corresponding to classical fields: if two classical fields are identical in a certain region, then their corresponding quantum states have the same reduced density matrix in that region. 
	This implies that, although the nonlocal nature of spatial entanglement seems to contradict the locality of classical fields, we can still obtain a consistent correspondence between classical fields and quantum states. And we only need to know the classical field in a local region to derive the corresponding reduced density matrix for that region. Thus, the correspondence between classical fields and quantum states is not only global but also local, which also avoids inconsistencies from arising. 
		While the spatial entanglement of quantum states does not undermine the consistency of the correspondence between quantum states and classical fields, it does lead to other counterintuitive conclusions. For instance, when two quantum states with the same reduced density matrix in a certain region are superposed, the resulting superposition state no longer preserves the original reduced density matrix in that region due to the spatial entanglement.

	\begin{acknowledgments}
	We thank Yida Li for their helpful discussions. We especially thank Wanqiang Liu for proposing the free coherent states in the $\phi$-representation. The work of Q. Wang was supported in part by the National Key Research and Development Program of China (Grant No. 2017YFA0402204).  
	\end{acknowledgments}
	
	\appendix
	
	\renewcommand{\appendixname}{Appendix}

\section{Calculation for $\mathcal{N}(t)$ and $f(x,t)$}

\label{sec:AppA}
In this section, we consider the evolution of a given quantum state in the Schrodinger picture. A general state that can revert to classical field theory looks like a coherent state with a Gaussian wave function\eqref{gauwave}. To continue our calculation, we first define two new kernel functions $\mathcal{E}_{-1}(\boldsymbol{x},\boldsymbol{y})$ and $\mathcal{E}_2(\boldsymbol{x},\boldsymbol{y})$
\begin{equation}	
	\mathcal{E}_{-1}(\boldsymbol{x},\boldsymbol{y})
	\equiv\int\frac{d^3p}{(2\pi)^3}\frac{1}{E_{\boldsymbol{p}}} e^{i\boldsymbol{p}\cdot(\boldsymbol{x}-\boldsymbol{y})}
	\;,
\end{equation}
\begin{equation}
	\label{prop2}	
	\mathcal{E}_{2}(\boldsymbol{y},\boldsymbol{z})
	\equiv\int d^3x\; \mathcal{E}(\boldsymbol{x},\boldsymbol{y}) \mathcal{E}(\boldsymbol{x},\boldsymbol{z})
	=\int\frac{d^3p}{(2\pi)^3} e^{i\boldsymbol{p}\cdot(\boldsymbol{y}-\boldsymbol{z})} E_{\boldsymbol{p}}^2
	\; .
\end{equation}
The kernel $\mathcal{E}_2$ acts on an arbitrary function $g(\boldsymbol{x})$

\begin{equation}	
	\label{ker2act}
	\int d^3x\;\mathcal{E}_{2}(\boldsymbol{x},\boldsymbol{y}) g(\boldsymbol{x})
	=-\nabla^2g(\boldsymbol{y})+m^2g(\boldsymbol{y})
	\; .
\end{equation}
The following are the relations between the three kernel functions $\mathcal{E}_{-1}$, $\mathcal{E}$ and $\mathcal{E}_2$
\begin{equation}\label{kernnor}	
	\int d^3x\;\mathcal{E}(\boldsymbol{x},\boldsymbol{y})\mathcal{E}_{-1}(\boldsymbol{x},\boldsymbol{z})
	=\delta(\boldsymbol{y}-\boldsymbol{z})
	\;,
\end{equation}
\begin{equation}
	\label{2-1}	
	\int d^3x\;\mathcal{E}_2(\boldsymbol{x},\boldsymbol{y})\mathcal{E}_{-1}(\boldsymbol{x},\boldsymbol{z})
	=\mathcal{E}(\boldsymbol{y},\boldsymbol{z})
	\; .
\end{equation}

It is convenient to calculate the Schrodinger equation in the $\phi$-representation. In this representation, the momentum operator acts as a gradient functional operator $\hat\pi(\boldsymbol{x})=-i\frac{\delta}{\delta \phi(\boldsymbol{x})}$. For the quantum state \eqref{gauwave} in this representation, we have

\begin{equation}
	\begin{split}	
		\label{hatpi}
		\hat\pi(\boldsymbol{x})\left<\phi|\varphi\right>
		&=i\int d^3y \;
		\mathcal{E}(\textbf x,\textbf y)[\phi(\textbf y)-f(\textbf y)]
		\left<\phi|\varphi\right>
		\; .
	\end{split}
\end{equation}
And then
\begin{equation}
	\begin{split}	
		\hat\pi^2(\boldsymbol{x})\left<\phi|\varphi\right>
		=\mathcal{E}(\boldsymbol{x},\boldsymbol{x})
		\left<\phi|\varphi\right>
		-\int d^3y \;
		\mathcal{E}(\boldsymbol{x},\boldsymbol{y})[\phi(\boldsymbol{y})-f(\boldsymbol{y})]
		\int d^3z\;
		\mathcal{E}(\boldsymbol{x},\boldsymbol{z})[\phi(\boldsymbol{z})-f(\boldsymbol{z})]
		\left<\phi|\varphi\right>
		 .
	\end{split}
\end{equation}
Considering the property $\eqref{prop2}$, we find that
\begin{equation}
	\int d^3x\;\hat\pi^2(\boldsymbol{x})\left<\phi|\varphi\right>
	=\int  d^3x\;\mathcal{E}(\boldsymbol{x},\boldsymbol{x})
	\left<\phi|\varphi\right>
	-\int d^3y  d^3z
	[\phi(\boldsymbol{y})-f(\boldsymbol{y})] [\phi(\boldsymbol{z})-f(\boldsymbol{z})]\mathcal{E}_2(\boldsymbol{y},\boldsymbol{z})
	\left<\phi|\varphi\right>
	.
\end{equation}

Then, we let the Hamiltonian act on a given quantum state \eqref{gauwave} and consider the property \eqref{ker2act} 
\begin{equation}
	\begin{split}
		\label{H}
		\hat H \left<\phi|\varphi\right>&=\int d^3x
		\left\{\left[\frac{1}{2}\hat\pi^2(\boldsymbol{x})+\frac{1}{2}\left(\nabla\hat\phi(\boldsymbol{x})\right)^2+\frac{1}{2}m_0^2\hat\phi^2(\boldsymbol{x})+g\hat\phi(\boldsymbol{x}) j(\boldsymbol{x},t)\right]+C\right\}\left<\phi|\varphi\right>\\
		&=\left[ -\frac{1}{2}\int d^3y \int d^3z\;
		f(\boldsymbol{y})f(\boldsymbol{z}) \mathcal{E}_2(\boldsymbol{y},\boldsymbol{z} )
		+\int d^3y \int d^3z\;
		f(\boldsymbol{y}) \phi(\boldsymbol{z})\mathcal{E}_2(\boldsymbol{y},\boldsymbol{z} )\right.\\
		&~~~~~~\left.+g\int  d^3x\; \phi(\boldsymbol{x}) j(\boldsymbol{x},t)
		+\frac{1}{2}\int  d^3x\; \mathcal{E}(\boldsymbol{x},\boldsymbol{x})+\Lambda \right]
		\left<\phi|\varphi\right>
		\; .
	\end{split}
\end{equation}
Let's recall that the wave function of the quantum state undergoing time evolution is assumed to have the following form (see \eqref{assumption})
\begin{equation}
	\left<\phi|\varphi(t)\right>=\mathcal{N}(t)\exp\left\{-\frac{1}{2}\int d^3xd^3y \;
	\mathcal{E}(\boldsymbol{x},\boldsymbol{y})[\phi(\boldsymbol{x})-f(\boldsymbol{x},t)][\phi(\boldsymbol{y})-f(\boldsymbol{y},t)]\right\}
	\; .
\end{equation}
The time-dependent normalization coefficient arises from the evolution of the wave function, which suggests that we can write the time-dependent coefficient as $\mathcal{N}(t)=\mathcal{N}_0e^{F(f,t)}$, where $F(f,t)$ depends on the center position function $f$ and time $t$. We can consider the Schrodinger equation $i\frac{d}{dt}\left<\phi|\varphi(t)\right>=\hat H \left<\phi|\varphi(t)\right>$ to calculate the explicit form of $f$ and $F$ in the quantum state. For the time derivative term, we have
\begin{equation}
	\begin{split}	
		\label{t}
		\frac{d}{dt}\left<\phi|\varphi(t)\right>
		&=\frac{d}{dt}
		\left[ \mathcal{N}_0e^{F(f,t)}
		\exp\left\{-\frac{1}{2}\int d^3xd^3y \;
		\mathcal{E}(\boldsymbol{x},\boldsymbol{y})[\phi(\boldsymbol{x})-f(\boldsymbol{x},t)][\phi(\boldsymbol{y})-f(\boldsymbol{y},t)]\right\}\right]\\
		&=\left\{\int d^3xd^3y \;
		\mathcal{E}(\boldsymbol{x},\boldsymbol{y})\dot f(\boldsymbol{x},t)[\phi(\boldsymbol{y})-f(\boldsymbol{y},t)]+\frac{\partial}{\partial t}F(f,t)  \right\}  \left<\phi|\varphi(t)\right>
		\; ,
	\end{split}
\end{equation}
where $\dot f(\boldsymbol{x},t)\equiv \frac{\partial}{\partial t} f(\boldsymbol{x},t)$. 
Hence, by considering both sides of the Schrödinger equation \eqref{H} and \eqref{t}, we can express the Schrödinger equation as follows
\begin{equation}
	\begin{split}
		\label{phi}	
		&i\left\{\int d^3xd^3y \;
		\mathcal{E}(\boldsymbol{x},\boldsymbol{y})\dot f(\boldsymbol{x},t)[\phi(\boldsymbol{y})-f(\boldsymbol{y},t)]+\frac{\partial}{\partial t}F(f,t) \right\}  \left<\phi|\varphi\right>\\
		&=\left[ -\frac{1}{2}\int d^3y \int d^3z\;
		f(\boldsymbol{y},t)f(\boldsymbol{z},t) \mathcal{E}_2(\boldsymbol{y},\boldsymbol{z} )
		+\int d^3y \int d^3z\;
		f(\boldsymbol{y},t) \phi(\boldsymbol{z})\mathcal{E}_2(\boldsymbol{y},\boldsymbol{z} )\right.\\
		&~~~~~\left.+g\int d^3x\; \phi(\boldsymbol{x}) j(\boldsymbol{x},t)
		+\frac{1}{2}\int  d^3x\;\mathcal{E}(\boldsymbol{x},\boldsymbol{x})+\Lambda \right]
		\left<\phi|\varphi\right>
		\;  .
	\end{split}
\end{equation}
The Schrodinger equation \eqref{phi} is valid for any arbitrary $\phi$, which require
\begin{equation}
	\label{ft}
	i\int d^3x\;
	\mathcal{E}(\boldsymbol{x},\boldsymbol{y})\dot f(\boldsymbol{x},t) 
	=\int d^3x \;
	f(\boldsymbol{x},t)\mathcal{E}_2(\boldsymbol{x},\boldsymbol{y} )
	+g j(\boldsymbol{y},t)
	\; ,
\end{equation}
\begin{equation}	
	\label{F} 
	i\int d^3xd^3y \;
	\mathcal{E}(\boldsymbol{x},\boldsymbol{y})\dot f(\boldsymbol{x},t)f(\boldsymbol{y},t)
	-i\frac{\partial}{\partial t}F(f,t)
	= \frac{1}{2}\int d^3y  d^3z\;
	f(\boldsymbol{y},t)f(\boldsymbol{z},t) \mathcal{E}_2(\boldsymbol{y},\boldsymbol{z} )
	\; ,
\end{equation}
\begin{equation}
	\Lambda=-\frac{1}{2}\int  d^3x\;\mathcal{E}(\boldsymbol{x},\boldsymbol{x})
	\; .
\end{equation}

Then, we separate $f$ into real and imaginary parts
\begin{equation}
	\label{f12}
	f(\boldsymbol{x},t)=f_1(\boldsymbol{x},t)+i\int d^3y\;\mathcal{E}_{-1}(\boldsymbol{x},\boldsymbol{y} )f_2(\boldsymbol{y},t)
	\; ,
\end{equation}
where $f_1$ and $f_2$ are both real. Substituting $\eqref{f12}$ into $\eqref{ft}$ and using \eqref{ker2act} , we obtain
\begin{equation}
	\label{id}
	i\int d^3x\;
	\mathcal{E}(\boldsymbol{x},\boldsymbol{y}) \dot f_1(\boldsymbol{x},t)
	-\dot f_2(\boldsymbol{y},t)
	=i \int d^3z\; \mathcal{E}(\boldsymbol{y},\boldsymbol{z})f_2(\boldsymbol{z},t) 
	-\nabla^2  f_1(\boldsymbol{y},t)+m_0^2 f_1(\boldsymbol{y},t)+g j(\boldsymbol{y},t)
	\; .
\end{equation}
Comparing the real and imaginary parts of both sides in \eqref{id}, we find the relation between the real part $f_1$ and imaginary part $f_2$
\begin{equation}
	\begin{split}	
		\label{f2}
		-\dot f_2(\boldsymbol{y},t)
		&=-\nabla^2  f_1(\boldsymbol{y},t)+m_0^2 f_1(\boldsymbol{y},t)+g j(\boldsymbol{y},t)\\
		\dot f_1(\boldsymbol{x},t)&=f_2(\boldsymbol{x},t) 
		\; .
	\end{split}
\end{equation}

On the other hand, from \eqref{F}, we can work out the explicit form of $F(f,t)$ in terms of the center function $f(\boldsymbol{x},t)$:
\begin{equation}
	\begin{split}
		\label{F(f)}
		F(f,t)
		=\frac{1}{2}\int d^3xd^3y \;
		\mathcal{E}(\boldsymbol{x},\boldsymbol{y}) f(\boldsymbol{x},t)f(\boldsymbol{y},t)
		+i\frac{1}{2}\int_{t_0}^{t}d\tau \int d^3xd^3y \;
		\mathcal{E}_2(\boldsymbol{x},\boldsymbol{y} )
		f(\boldsymbol{x},\tau)f(\boldsymbol{y},\tau) 
		\; ,
	\end{split}
\end{equation}
where $t_0$ is an arbitrary initial time. Fianlly, the normalization coefficient $\mathcal{N}(t)=\mathcal{N}_0e^{F(f,t)}$ can be expressed as
\begin{equation}
	\label{Fnew}	
	\begin{split}
	\mathcal{N}(t)=\mathcal{N}_0\exp\Bigg\{\frac{1}{2}&\int d^3xd^3y \;
		\mathcal{E}(\boldsymbol{x},\boldsymbol{y}) f(\boldsymbol{x},t)f(\boldsymbol{y},t)
		\\
		&+i\frac{1}{2}\int_{t_0}^{t}d\tau\int d^3xd^3y \;
		\mathcal{E}_2(\boldsymbol{x},\boldsymbol{y} )
		f(\boldsymbol{x},\tau)f(\boldsymbol{y},\tau) \Bigg\}
	\; .	
	\end{split}
\end{equation}

\section{Wave function in the $\pi$-representation}

\label{sec:AppB}
To demonstrate whether the wave function of the quantum state is localized around $\pi_{\textcolor{black}{\text{class}}}$, we can calculate the Gaussian wave function in the $\pi$-representation. The eigenstate and eigenvalue of the conjugate momentum operator $\hat{\pi}(\boldsymbol{x})$ are $\left|\pi\right>$ and $\pi(\boldsymbol{x})$, which means $\hat\pi(\boldsymbol{x})\left|\pi\right>=\pi(\boldsymbol{x})\left|\pi\right>$. The commutation relation of $\hat\pi(\boldsymbol{x})$ and  $\hat\phi(\boldsymbol{x})$ yields
\begin{equation}	
	\left<\phi|\pi\right>=C\exp{\left\{i\int d^3x\pi(\boldsymbol{x})\phi(\boldsymbol{x})\right\} }				
	\; .
\end{equation}
Using completeness, we can write the wave function for state $\left|\varphi(t)\right>$ in the $\pi$-representation
\begin{equation}
	\begin{split}	
		\label{pi function}
	&	\left<\pi|\varphi(t)\right>\\
	&=\int \mathcal{D}\phi \left<\pi|\phi\right>\left<\phi|\varphi(t)\right>\\
		&=\mathcal{N'}(t)C\exp\left\{i\int d^3x\;[\pi_{\textcolor{black}{\text{class}}}(\boldsymbol{x},t)-\pi(\boldsymbol{x})]\phi_{\textcolor{black}{\text{class}}}(\boldsymbol{x},t)\right\}\\
		&~~~~\times \int \mathcal{D}\phi \exp\left\{
		-\frac{1}{2}\int d^3xd^3y \;
		\mathcal{E}(\boldsymbol{x},\boldsymbol{y})
		\phi(\boldsymbol{x})\phi(\boldsymbol{y})
		+i\int d^3x\; [\pi_{\textcolor{black}{\text{class}}}(\boldsymbol{x},t)-\pi(\boldsymbol{x})]\phi(\boldsymbol{x})\right\}
		\; .
	\end{split}
\end{equation}
To finish the integral, we define a matrix $T(\boldsymbol{x},\boldsymbol{x}')$ to diagonalize the matrix $\mathcal{E}(\boldsymbol{x},\boldsymbol{y})$
\begin{equation}
	\int d^3xd^3y \;
	T(\boldsymbol{x},\boldsymbol{x}')\mathcal{E}(\boldsymbol{x},\boldsymbol{y}) T(\boldsymbol{y},\boldsymbol{y}')=\delta(\boldsymbol{x}'-\boldsymbol{y}')
	\; .
\end{equation}
Considering the normalization condition \eqref{kernnor} for the kernel function $\mathcal{E}(\boldsymbol{x}.\boldsymbol{y})$, we can explicitly write the auxiliary matrix $T(\boldsymbol{x},\boldsymbol{x}')$
\begin{equation}
	T(\boldsymbol{x},\boldsymbol{x}')
	\equiv\int\frac{d^3p}{(2\pi)^3}\frac{1}{\sqrt{E_{\boldsymbol{p}}}} e^{i\boldsymbol{p}\cdot(\boldsymbol{x}-\boldsymbol{y})}
	\; .
\end{equation}
To diagonalize with the auxiliary matrix $T(\boldsymbol{x},\boldsymbol{x}')$, we need to change the integration variable $\phi$ to $\phi'$ via
\begin{equation}	
	\phi(\boldsymbol{x})=\int d^3x'\; T(\boldsymbol{x},\boldsymbol{x}') \phi'(\boldsymbol{x}')
	\; ,
\end{equation}
and change the measure of the integral to
\begin{equation}
	\mathcal{D}\phi(\boldsymbol{x})= J \mathcal{D}\phi'(\boldsymbol{x}')
	\; ,
\end{equation}
where $J=\left|\text{det}[T(\boldsymbol{x},\boldsymbol{x}')]\right|$ is the determinant of the auxiliary matrix. Then, the integral in $\eqref{pi function}$ can be done
\begin{equation}
	\begin{split}	
		\label{int}
		&\int \mathcal{D}\phi \exp\left\{
		-\frac{1}{2}\int d^3xd^3y \;
		\mathcal{E}(\boldsymbol{x},\boldsymbol{y})
		\phi(\boldsymbol{x})\phi(\boldsymbol{y})
		+i\int d^3x\;[\pi_{\textcolor{black}{\text{class}}}(\boldsymbol{x},t)-\pi(\boldsymbol{x})]\phi(\boldsymbol{x})\right\}\\
		&= J\int \mathcal{D}\phi' \exp\Bigg\{
		-\frac{1}{2}\int d^3xd^3y \;
		\mathcal{E}(\boldsymbol{x},\boldsymbol{y})
		\int d^3x'\; T(\boldsymbol{x},\boldsymbol{x}') \phi'(\boldsymbol{x}')
		\int d^3y'\; T(\boldsymbol{y},\boldsymbol{y}') \phi'(\boldsymbol{y}')\\
		&~~~~~~~~~~~~~~~~~~~~~~~+i\int d^3x\;[\pi_{\textcolor{black}{\text{class}}}(\boldsymbol{x},t)-\pi(\boldsymbol{x})]
		\int d^3x'\; T(\boldsymbol{x},\boldsymbol{x}')\phi(\boldsymbol{x}')\Bigg\}\\
		&=J\int \mathcal{D}\phi' \exp\Bigg\{
		-\frac{1}{2}\int d^3x'\;	\phi'(\boldsymbol{x}') \phi'(\boldsymbol{x}')\\
		&~~~~~~~~~~~~~~~~~~~~~~~~+i\int d^3x'\int d^3x\; T(\boldsymbol{x},\boldsymbol{x}')[\pi_{\textcolor{black}{\text{class}}}(\boldsymbol{x},t)-\pi(\boldsymbol{x})]\phi(\boldsymbol{x}')\Bigg\}\\
		&=J\prod\limits_{x'} \int_{-\infty}^{\infty} d\phi'(\boldsymbol{x}') \exp\Bigg\{
		-\frac{1}{2} d^3x'\;	\phi'(\boldsymbol{x}') \phi'(\boldsymbol{x}')\\
		&~~~~~~~~~~~~~~~~~~~~~~~~~~~~~~~~~~+i d^3x'\int d^3x\; T(\boldsymbol{x},\boldsymbol{x}')[\pi_{\textcolor{black}{\text{class}}}(\boldsymbol{x},t)-\pi(\boldsymbol{x})]\phi(\boldsymbol{x}')\Bigg\}\\
		&=J\prod\limits_{x'} 2\int_0^\infty d\phi'(\boldsymbol{x}') 
		\exp{\left[	-\frac{1}{2} d^3x'\;	\phi'(\boldsymbol{x}') \phi'(\boldsymbol{x}')\right]}\\
		&~~~~~~~~~~~~~~~~~~~~~~~~~~\times \cos\left\{ d^3x'\int d^3x\; T(\boldsymbol{x},\boldsymbol{x}')[\pi_{\textcolor{black}{\text{class}}}(\boldsymbol{x},t)-\pi(\boldsymbol{x})]\phi(\boldsymbol{x}')\right\}\\
		&=J\prod\limits_{x'} \left( \frac{\sqrt\pi}{\sqrt{\frac{1}{2} d^3x'}}\right)
		\exp\Bigg\{-\frac{1}{2}\int d^3x'\int d^3x\; T(\boldsymbol{x},\boldsymbol{x}')[\pi_{\textcolor{black}{\text{class}}}(\boldsymbol{x},t)-\pi(\boldsymbol{x})]\\
		&~~~~~~~~~~~~~~~~~~~~~~~~~~~~~~~~~~~~~~~~~~~~~~~~~~~~\times\int d^3y\; T(\boldsymbol{y},\boldsymbol{x}')[\pi_{\textcolor{black}{\text{class}}}(\boldsymbol{y},t)-\pi(\boldsymbol{y})] \Bigg\}.
	\end{split}
\end{equation}
Define a constant $C'=\prod\limits_{x'} \left(\sqrt\pi/\sqrt{\frac{1}{2} d^3x'} \right)$. Furthermore, substituting $\eqref{int}$ into $\eqref{pi function}$, we obtain
\begin{equation}\label{TTT}
	\begin{split}	
		&\left<\pi|\varphi(t)\right>\\
  &=J\mathcal{N'}(t)CC'
		\exp\left\{i\int d^3x[\pi_{\textcolor{black}{\text{class}}}(\boldsymbol{x},t)-\pi(\boldsymbol{x})]\phi_{\textcolor{black}{\text{class}}}(\boldsymbol{x},t)\right\}\\
		&\times	  
		\exp\Bigg\{-\frac{1}{2}\int d^3x'\int d^3x\; T(\boldsymbol{x},\boldsymbol{x}')[\pi_{\textcolor{black}{\text{class}}}(\boldsymbol{x},t)-\pi(\boldsymbol{x})]
		\int d^3y\; T(\boldsymbol{y},\boldsymbol{x}')[\pi_{\textcolor{black}{\text{class}}}(\boldsymbol{y},t)-\pi(\boldsymbol{y})] \Bigg\}\\
		&=J\mathcal{N''}(t)C''
		\exp\Bigg\{-i\int d^3x\; \pi(\boldsymbol{x})\phi_{\textcolor{black}{\text{class}}}(\boldsymbol{x},t)\Bigg\}\\
		&\times	
		\exp\Bigg\{-\frac{1}{2}\int d^3x d^3y \Big(\int d^3x'\;T(\boldsymbol{x},\boldsymbol{x}')T(\boldsymbol{y},\boldsymbol{x}')\Big)
		[\pi_{\textcolor{black}{\text{class}}}(\boldsymbol{x},t)-\pi(\boldsymbol{x})][\pi_{\textcolor{black}{\text{class}}}(\boldsymbol{y},t)-\pi(\boldsymbol{y})] \Bigg\}
		,
	\end{split}
\end{equation}
where $\mathcal{N''}=\mathcal{N'}(t)\exp{\left[i\int d^3x\;\pi_{\textcolor{black}{\text{class}}}(\boldsymbol{x},t)\phi_{\textcolor{black}{\text{class}}}(\boldsymbol{x},t)\right]}$ is a new normalization coefficient.
Note that the integral of the auxiliary matrix in \eqref{TTT} can be simplified further in terms of the kernel function $\mathcal{E}_{-1}(\boldsymbol{x},\boldsymbol{y})$
\begin{equation}
	\int d^3x'\;T(\boldsymbol{x},\boldsymbol{x}')T(\boldsymbol{y},\boldsymbol{x}')=\int\frac{d^3p}{(2\pi)^3}\frac{1}{E_{\boldsymbol{p}}} e^{i\boldsymbol{p}\cdot(\boldsymbol{x}-\boldsymbol{y})}
	=\mathcal{E}_{-1}(\boldsymbol{x},\boldsymbol{y})
	\; .
\end{equation}
Then, the final expression of the wave function in the $\pi$-representation is
\begin{equation}\label{pi}
	\begin{split}
		\left<\pi|\varphi(t)\right>=J\mathcal{N''}(t)&C''
		\exp\left\{-i\int d^3x\;\pi(\boldsymbol{x})\phi_{\textcolor{black}{\text{class}}}(\boldsymbol{x},t)\right\}\\
		\times
		&\exp\left\{-\frac{1}{2}\int d^3x d^3y \; \mathcal{E}_{-1}(\boldsymbol{x},\boldsymbol{y})
		[\pi(\boldsymbol{x})-\pi_{\textcolor{black}{\text{class}}}(\boldsymbol{x},t)][\pi(\boldsymbol{y})-\pi_{\textcolor{black}{\text{class}}}(\boldsymbol{y},t)] \right\}	
		.
	\end{split}
\end{equation}
Therefore, the wave function is a Gaussian wave packet with the center located at $\pi_{\text{class}}$.

	\bibliographystyle{JHEP}
	\bibliography{Yukawa}

\end{document}